\documentclass[fleqn,aps,prb,twocolumn,showpacs,floatfix,longbibliography]{revtex4-2}
\usepackage{amsmath,amssymb,amsfonts,float,graphics,epsfig,epstopdf,color,verbatim,tabularx,bm,multirow,appendix,hyperref}

\usepackage{amsmath}
\usepackage{amssymb}
\usepackage{mathtools}
\usepackage{graphicx}
\usepackage{lipsum}
\usepackage{bm}
\usepackage{hyperref}
\usepackage{url}
\usepackage[utf8]{inputenc}
\usepackage{subfigure}
\usepackage{slashed,bbm}
\usepackage{graphics,psfrag,epsfig}
\usepackage{dsfont}
\usepackage{setspace}
\usepackage{wasysym}
\usepackage{slashed}
\usepackage{lipsum}
\usepackage{braket}
\usepackage{physics}
\usepackage{enumerate}

\usepackage{xcolor}

\definecolor{dark-red}{rgb}{0.4,0.15,0.15}
\definecolor{dark-blue}{rgb}{0.15,0.15,0.4}
\definecolor{medium-blue}{rgb}{0,0,0.5}
\hypersetup{
colorlinks, linkcolor={dark-blue},
citecolor={dark-blue}, urlcolor={medium-blue}
}
\usepackage{ulem}

\newcommand{\be}{\begin{equation}}
\newcommand{\ee}{\end{equation}}
\newcommand{\bea}{\begin{eqnarray}}
\newcommand{\eea}{\end{eqnarray}}

\begin{document}

\title{Separability transitions in topological states induced by local decoherence}

\author{Yu-Hsueh Chen}
\affiliation{Department of Physics, University of California at San Diego, La Jolla, California 92093, USA}
\author{Tarun Grover}
\affiliation{Department of Physics, University of California at San Diego, La Jolla, California 92093, USA}


\begin{abstract}

\noindent
We study states with intrinsic topological order subjected to local decoherence from the perspective of \textit{separability}, i.e., whether a decohered mixed state can be expressed as an ensemble of short-range entangled (SRE) pure states. We focus on toric codes and the X-cube fracton state and provide evidence for the existence of decoherence-induced separability transitions that precisely coincide with the threshold for the feasibility of active error correction. A key insight is that local decoherence acting on the `parent' cluster states of these models results in a Gibbs state. As an example, for the 2d (3d) toric code subjected to bit-flip errors, we show that the decohered density matrix can be written as a convex sum of SRE states for $p > p_c$, where $p_c$ is related to the paramagnetic-ferromagnetic transition in the 2d (3d) random-field bond Ising model along the Nishimori line.
\end{abstract}

\maketitle

In this work we will explore aspects of many-body topological states subjected to decoherence from the perspective of \textit{separability}, i.e., whether the resulting mixed state can be expressed as a convex sum of short-range entangled (SRE) states \cite{werner1989,hastings2011topological,horodecki2009quantum}. This criteria is central to the definition of what constitutes an SRE or long-range entangled (LRE) mixed state, and various measures of mixed-state entanglement, such as negativity\cite{peres1996, horodecki1996, eisert99, vidal2002, plenio2005logarithmic, horodecki2009quantum} and entanglement of formation \cite{bennett1996}, are defined so as to quantify non-separability. We will be particularly interested in decoherence-induced ``separability transitions'', i.e., transitions tuned by decoherence such that the density matrix in one regime is expressible as a convex sum of SRE states, and in the other regime, it is not. One salient distinction between pure state versus mixed-state dynamics is that although a short-depth unitary evolution cannot change long-range entanglement encoded in a pure state, a short-depth local \textit{channel} can fundamentally alter long-range mixed-state entanglement. 
Therefore,  even the limited class of mixed states that are obtained by the action of local short-depth channels on an entangled pure state offer an opportunity to explore mixed-state phases and phase transitions \cite{de2022symmetry,ma2022average,lee2022symmetry,lee2023quantum, fan2023diagnostics,bao2023mixed,zou2023channeling,lu2023mixed,ma2023topological,ma2023exploring,su2023conformal,wang2023intrinsic,sang2023mixed}. We will  focus on mixed states that are obtained via subjecting several well-understood topologically ordered phases of matter to short-depth quantum channels.

\begin{figure}[t]
	\centering
	\includegraphics[width=0.45\textwidth]{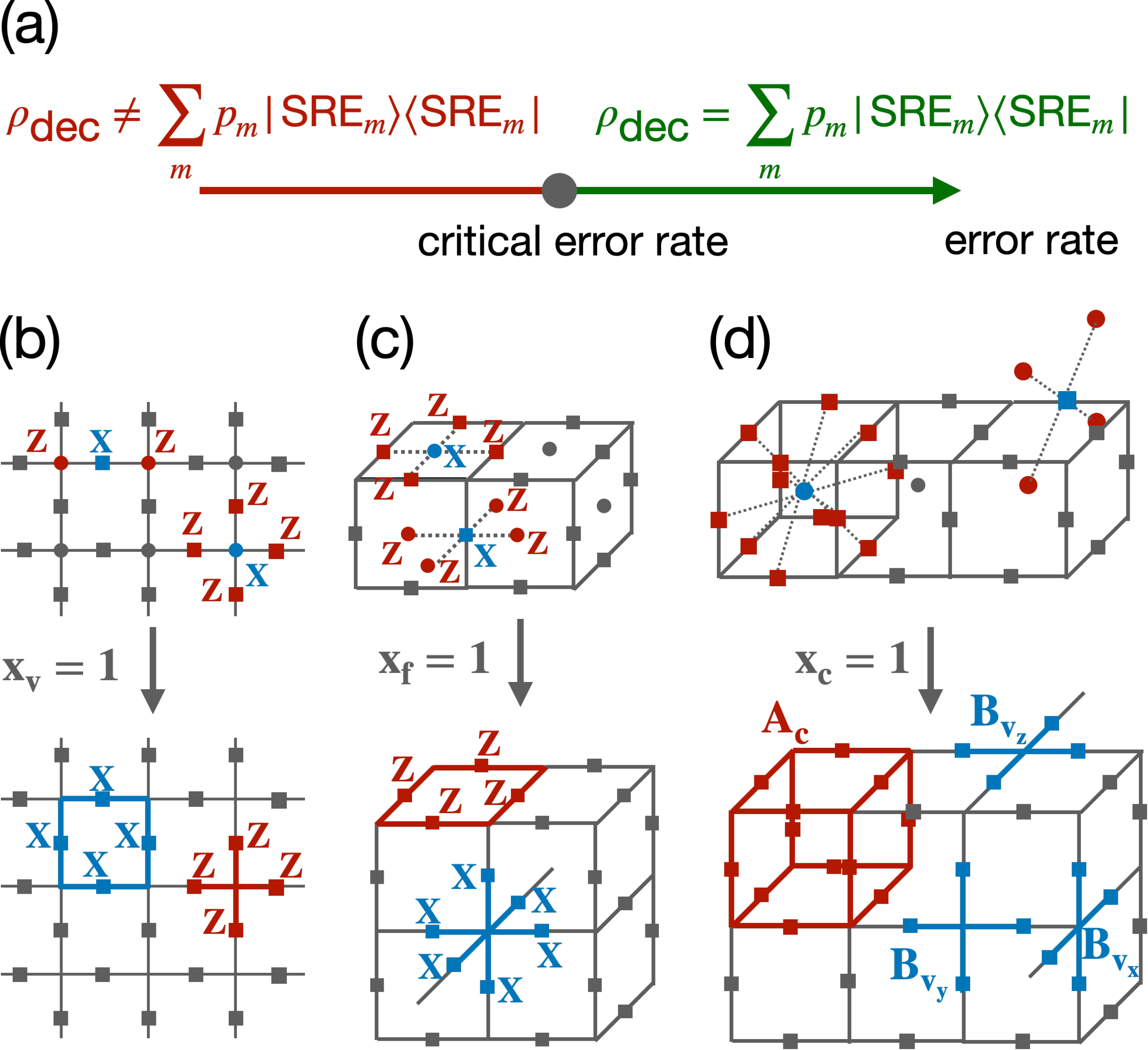}
	\caption{ 
		(a)  Topological orders under local decoherence can undergo a separability transition, where only above a certain critical error rate, the decohered mixed state $\rho_{\textrm{dec}}$ can be written as a convex sum of SRE pure states.
		The bottom depicts the parent cluster states and their offspring models obtained by appropriate measurements (indicated by an arrow) (b) 2d cluster Hamiltonian and 2d toric code, (c) 3d cluster Hamiltonian and 3d toric code, and (d) ``Cluster-X'' Hamiltonian \cite{verresen2021efficiently} and the X-cube Hamiltonian.} 
	\label{Fig:toric_xcube_main}
\end{figure}

Error-threshold theorems \cite{shor1996fault,aharonov1997fault,kitaev2003fault,knill1998resilient,preskill1998reliable,terhal2015quantum} suggest a topologically ordered pure state is perturbatively stable against decoherence from a short-depth, local quantum channel, leading to the possibility of a phase transition as a function of the decoherence rate \cite{aharonov2000quantum}. Such transitions were originally studied  from the perspective of quantum error correction (QEC) in Refs.\cite{dennis2002,wang2003confinement} and more recently using mixed-state entanglement measures such as topological negativity \cite{fan2023diagnostics}, and other non-linear functions of the density matrix (Refs.\cite{fan2023diagnostics, bao2023mixed, lee2023quantum}). These approaches clearly establish at least two different mixed-state phases: one where the topological qubit can be decoded, and the other where it can't. However, it is not obvious if the density matrix in the regime where decoding fails can be expressed as a convex sum of SRE pure states, which, following Refs.\cite{werner1989,hastings2011topological}, we will take as the definition of an SRE mixed state. Our main result is that for several topologically ordered phases subjected to local decoherence, which are relevant for quantum computing \cite{dennis2002,wang2003confinement,fowler2012surface},  one can explicitly write down the decohered mixed state as a convex sum of pure states which we argue all undergo a topological phase transition, from being long-ranged entangled to short-ranged entangled, at a threshold that precisely corresponds to the optimal threshold for QEC. We find that the universality class of such a separability transition also coincides with that corresponding to the QEC error-recovery transition. Therefore, in these examples, we argue that the error-recovery transition does indeed coincide with a many-body separability transition. As discussed below, our method also provides a new route to obtain the statistical mechanics models relevant for the quantum error-correcting codes \cite{nishimori1981internal,dennis2002,wang2003confinement,fan2023diagnostics}. 



Let us begin by considering the ground state of the 2d toric code  (see Fig.\ref{Fig:toric_xcube_main}(b)) with Hamiltonian $	H_{\textrm{2d toric}} = - \sum_v (\prod_{e \in v} Z_e) - \sum_{p} (\prod_{e \in p } X_e)$ subjected to phase-flip errors. 
The Hilbert space consists of qubits residing on the edges (denoted as `$e$') of a square lattice and we assume periodic boundary conditions. 
Denoting the ground state as $\rho_0$,  the Kraus map corresponding to the phase-flip errors act on an edge $e$ as: $\mathcal{E}_e[\rho_0] = p Z_e \rho_0 Z_e + (1-p) \rho_0$, and the full map is given by the composition of this map over all edges.
The key first step is to utilize the idea of duality \cite{kramers1941statistics,wegner1971duality, williamson2016fractal, kubica2018ungauging,rakovszky2023physics} by identifying the corresponding `parent' cluster Hamiltonian (in the sense of Refs.\cite{raussendorf2005long,aguado2008creation,verresen2021efficiently,tantivasadakarn2021long}). Interestingly, the application of the aforementioned Kraus map to its ground state results in a Gibbs state. For the problem at hand, consider
$H_{\textrm{2d cluster}} = \sum_v h_{v} + \sum_e h_{e}$
where  $h_v = - X_v (\prod_{e \ni v} Z_e)$ and $h_e = - X_e (\prod_{v \in e} Z_v)$ whose Hilbert space consists of qubits both on the vertices and the edges of the square lattice (Fig.\ref{Fig:toric_xcube_main}(b)). The ground state density matrix $\rho_{0}$ of the 2d toric code can be written as  $\rho_{0} \propto \langle x_\mathbf{v} = 1| \rho_{C,0} |x_\mathbf{v} = 1\rangle$,
where $|x_\mathbf{v} = 1\rangle = \otimes_v |x_v = 1\rangle$ is the product state in the Pauli-$X$ basis, and $\rho_{C,0} \,(\propto \prod_e (I - h_e) \prod_v (I-h_v))$ is the ground state of $H_{\textrm{2d cluster}} $. The projection selects one specific ground state of the toric code that is an eigenvector of the non-contractible Wilson loops $W_{\ell} = \prod_{e \in \mathcal{\ell}} X_e$ with eigenvalue +1 along both cycles $\mathcal{\ell}$ of the torus. A simple calculation shows that $\mathcal{E}_e[\rho_{C,0}] \propto e^{-\beta \sum_e h_{e}}\prod_v (I -h_v)$ where $\tanh(\beta) = 1-2p$. This implies that the decohered density matrix $\rho$ of the toric code is $\rho 
 \propto \langle  x_\mathbf{v} = 1|
e^{-\beta \sum_e h_{e}}	| x_\mathbf{v} = 1\rangle P_{Z}$
where $P_Z = \prod_v (I+ \prod_{e \ni v} Z_e)$. 
By inserting a complete set of states, one may simplify the above expression as $\rho \propto P_{Z} \rho_e P_{Z}$ where $\rho_e =  \sum_{x_\mathbf{e}}  \mathcal{Z}_{\text{2d Ising}, x_\mathbf{e}}  |x_\mathbf{e}\rangle \langle x_\mathbf{e}|$ and $\mathcal{Z}_{\text{2d Ising}, x_\mathbf{e}}  = \sum_{z_\mathbf{v}} e^{\beta \sum_{e}  x_{e} \prod_{v \in e } z_v} $  is the partition function of the 2d Ising model with Ising interactions determined by $\{ x_e \}$. 
Thus, $\rho \propto \sum_{x_\mathbf{e}}   \mathcal{Z}_{\text{2d Ising}, x_\mathbf{e}}  |\Omega_{x_\mathbf{e}} \rangle \langle \Omega_{x_\mathbf{e}} | $, where $|\Omega_{x_\mathbf{e}} \rangle \propto \prod_v (I+ \prod_{e \ni v}Z_e)|x_\mathbf{e}\rangle$ are nothing but a subset of toric code eigenstates. Note that in this derivation, the 2d Ising model emerges due to the $h_e$ terms in the parent cluster Hamiltonian, and ultimately, this will lead to the relation between the separability transition and the statistical mechanics of the 2d random-bond Ising model (RBIM) that also describes the error-recovery transition \cite{dennis2002}. We note that the above spectral representation of $\rho$ in terms of toric code eigenstates has also previously appeared in Ref.\cite{lee2023quantum}, using a different derivation.  Since non-contractible cycles of the torus will play an important role below, let us note that distinct eigenstates $|\Omega_{x_\mathbf{e}}\rangle $ can be uniquely specified by two labels: the first label corresponds to the set of local $\mathbb{Z}_2$  fluxes $f_p = \prod_{e \in p} x_e$ through  elementary plaquettes $p$, while the second label $\mathbf{L} = (L_x = \pm 1, L_y = \pm 1)$  with $L_x = \prod_{e \in \ell, e \parallel \hat{x}} x_e, L_y = \prod_{e \in \ell, e \parallel \hat{y}} x_e$ and $\ell$ a non-contractible loop along $\hat{x}/\hat{y}$ direction, specifies the topological sector (`\textbf{L}ogical data') in which $|\Omega_{x_\mathbf{e}}\rangle$ lives. 

We now probe the mixed state $\rho$ using the separability criteria, i.e., we ask whether it can be decomposed as a convex sum of SRE states. Clearly, the aforementioned spectral representation is not a useful decomposition since it involves toric code eigenstates which are all LRE. Taking cue from the argument for separability of the Gibbs state of toric codes \cite{lu2020detecting}, we decompose $\rho$ as $\rho = \sum_{z_\mathbf{e}} \rho^{1/2}|z_\mathbf{e}\rangle \langle z_\mathbf{e}| \rho^{1/2} \equiv \sum_m |\psi_m\rangle \langle \psi_m|$ where $\{z_\mathbf{e}\}$ are a  complete set of product states in the Pauli-$Z$ basis, and $|\psi_m\rangle = \rho^{1/2}|z_\mathbf{e}\rangle$. Generically, to determine whether $\rho$ is an SRE mixed state, one needs to determine whether \textit{each} $|\psi_m\rangle$ is SRE. However, for the current case of interest, it suffices to consider only $|\psi\rangle = \rho^{1/2}|m_0 \rangle$ with $|m_0 \rangle = |z_\mathbf{e} = 1\rangle$. The reason is as follows. The Gauss's law ($\prod_{e \ni v} Z_e = 1$) implies that the Hilbert space only contains states that are closed loops in the $Z$ basis. Therefore, one may write $|m\rangle = g_x |m_0 \rangle$ where $g_x$  is a product of \textit{single-site} Pauli-$X$s forming closed loops. Since  $[g_x, \rho] = 0$, this implies that $|\psi^{}_m\rangle \equiv  |\psi_{g_x}\rangle = g_x |\psi\rangle$, and therefore, if $|\psi\rangle$ is SRE (LRE), so is $|\psi_{g_x}\rangle$. $\rho(\beta)$ may then be written as: $\rho(\beta) =  \sum_{g_x} |\psi_{g_x}(\beta)\rangle \langle \psi_{g_x}(\beta)|$. Now, using the aforementioned spectral representation of $\rho$, the (non-normalized) state $|\psi^{} \rangle = \rho^{1/2}|z_\mathbf{e} = 1\rangle$ is:

\begin{align}
\label{Eq:2d_toricCDA_Q}
|\psi^{} (\beta) \rangle & \propto  \sum_{x_\mathbf{e}} [\mathcal{Z}_{\text{2d Ising}, x_\mathbf{e}} (\beta)]^{1/2} |x_\mathbf{e} \rangle,
\end{align}
It is easy to see that when $\beta = \infty$, $|\psi\rangle \propto |\Omega_{0} \rangle$, the non-decohered toric code ground state, while when $\beta = 0$, $|\psi\rangle \propto |z_\mathbf{e} = 1\rangle$ is a product state. This suggests a phase transition for $|\psi(\beta)\rangle$ from being an LRE state to an SRE state as we increase the error rate $p$ (i.e. decrease $\beta$). We will now show that this is indeed the case.

We first consider the expectation value  of the `anyon condensation operator' (also known as `t Hooft loop) in state $|\psi(\beta)\rangle$, defined as \cite{hooft1978phase,kogut1979introduction, fradkin2013field,bao2023mixed} $T_{\mathcal{\tilde{\ell}}} = \prod_{e \in \mathcal{\tilde{\ell}}} Z_e$, where $\mathcal{\tilde{\ell}}$  denotes a homologically non-contractible loop on the dual lattice (in the language of $\mathbb{Z}_2$ gauge theory \cite{kogut1975hamiltonian,fradkin2013field}, $Z_e \sim e^{i \pi \textrm{(Electric field)}_e}$). 
Physically, $\langle T_{\mathcal{\tilde{\ell}}}  \rangle \equiv \langle \psi|T_{\mathcal{\tilde{\ell}}}|\psi\rangle/\langle \psi|\psi\rangle$ is the amplitude of tunneling from one logical subspace to an orthogonal one, and therefore it is zero in the $\mathbb{Z}_2$ topologically ordered phase, and non-zero in the topologically trivial phase (=anyon condensed phase) \footnote{We note that the `t Hooft loop works as an order parameter for detecting topological order in our pure state $|\psi\rangle$ because the Gauss's law ($\prod_{e \ni v} Z_e = 1 $) is satisfied exactly and therefore there are no dynamical charges.  Besides, note that $\tr(\rho T_{\mathcal{\tilde{\ell}}} )$, where $\rho$ is the decohered mixed state, is identically zero, and clearly does not capture mixed-state entanglement.}. Indeed, one may easily verify that $\langle T_{\mathcal{\tilde{\ell}}} \rangle = 0 \,(1)$ when $\beta= \infty\, (\beta = 0)$. Using Eq.\eqref{Eq:2d_toricCDA_Q},  $T_{\mathcal{\tilde{\ell}}}$ flips spins along the curve $\mathcal{\tilde{\ell}}$ (i.e., $x_e \rightarrow - x_e, \forall e \in \mathcal{\tilde{\ell}}$), and we denote the corresponding configuration as $x_{\mathcal{\tilde{\ell}}, \mathbf{e}}$. While $x_{\mathcal{\tilde{\ell}}, \mathbf{e}}$ and $x_{ \mathbf{e}}$ have the same flux through every elementary plaquette, they live in different logical  sectors $\mathbf{L}$. Therefore, $T_{\mathcal{\tilde{\ell}}} |\psi\rangle \propto \sum_{x_\mathbf{e}} [\mathcal{Z}_{ x_\mathbf{e}}]^{1/2} |x_{\mathcal{\tilde{\ell}}, \mathbf{e}}\rangle = \sum_{x_\mathbf{e}} [\mathcal{Z}_{ x_{\mathcal{\tilde{\ell}}, \mathbf{e}}}]^{1/2}|x_\mathbf{e}\rangle $, where we have suppressed the subscript `2d Ising' under the partition function $\mathcal{Z}$ for notational convenience. Thus,
\begin{equation}
\label{Eq:noncont_wilson2d}
\begin{aligned}
\langle T_{\mathcal{\tilde{\ell}}} \rangle & = \frac{\sum_{x_\mathbf{e}} \sqrt{\mathcal{Z}_{ x_\mathbf{e}} \mathcal{Z}_{x_{\mathcal{\tilde{\ell}}, \mathbf{e}}} } }{\sum_{x_\mathbf{e}} \mathcal{Z}_{ x_\mathbf{e}}} 
= \frac{\sum_{x_\mathbf{e}} \mathcal{Z}_{x_\mathbf{e}} e^{-\Delta F_{ x_{\mathcal{\tilde{\ell}},\mathbf{e}}}/2}}{\sum_{x_\mathbf{e}} \mathcal{Z}_{ x_\mathbf{e}}}  \\
& =\langle e^{-\Delta F_{\mathcal{\tilde{\ell}}}/2}\rangle \geq e^{-\langle \Delta F_{\mathcal{\tilde{\ell}}} \rangle /2}
\end{aligned}
\end{equation}
where $\Delta F_{ x_{\mathcal{\tilde{\ell}},\mathbf{e}}} =  - \log(\mathcal{Z}_{x_{\mathcal{\tilde{\ell}}, \mathbf{e}}} /\mathcal{Z}_{ x_\mathbf{e}})$ is the free energy cost of inserting a domain wall of size $|\tilde{\ell}| \sim L$ (= system's linear size) in the RBIM along the Nishimori line \cite{dennis2002}, and we have used Jensen's inequality in the last sentence. We are along the Nishimori line because the probability of a given gauge invariant label $\{f_\mathbf{p}, \mathbf{L}\}$ along the Nishimori line is precisely the partition function  $\mathcal{Z}_{x_\mathbf{e}}$ \cite{dennis2002}.
Since $\langle \Delta F_{\mathcal{\tilde{\ell}}} \rangle$, the disorder-averaged free energy cost, diverges with $L$ in the ferromagnetic phase of the RBIM while converges to a constant in the paramagnetic phase \cite{dennis2002}, Eq.\eqref{Eq:noncont_wilson2d} rigorously shows that for $p > p_c = p_{\text{2d RBIM}} \approx 0.109$ \cite{honecker2001universality}, $\langle T_{\mathcal{\tilde{\ell}}} \rangle $ saturates to a non-zero constant.
Therefore, $|\psi\rangle$ is a topologically trivial state when $p > p_c$, and hence the mixed state is SRE for $p > p_c$. In contrast, for $p < p_c$, due to non-vanishing ferromagnetic order (and associated domain wall cost) of the RBIM, we expect that $\langle T_{\mathcal{\tilde{\ell}}} \rangle  \sim e^{-\langle \Delta F_{\mathcal{\tilde{\ell}}} \rangle /2} \sim e^{- c L} \rightarrow 0$ in the thermodynamic limit  ($c > 0$ is a constant), implying that $|\psi\rangle$ is topologically ordered. This doesn't necessarily imply that the decohered state $\rho$ is long-range entangled for $p < p_c$, because there may exist some other way to decompose it as a sum of SRE pure states. However, for $p < p_c$, long-range entanglement as quantified by topological entanglement negativity is non-zero as shown in Ref.\cite{fan2023diagnostics}. Since a state of a form $\rho = \sum_i p_i |\textrm{SRE}\rangle_i \,{}_i\langle \textrm{SRE}|$ can be prepared by an ensemble of finite-depth unitary circuits starting with a product state, it is reasonable to assume that it cannot support long-range entanglement as quantified by any valid measure of entanglement. Assuming that topological negativity is one such measure (as supported by previous works \cite{lu2020detecting,fan2023diagnostics}), the above discussion implies that the state of our interest undergoes a separability transition at $p = p_c$.


Another diagnostic of topological order in pure states is the  (Renyi) topological  entanglement entropy (TEE) \cite{levin2006detecting,Kitaev06_1,flammia2009topological}. Dividing the system in real space as  $A \cup B$, and defining the reduced density matrix $\rho_A = \tr_{B} |\psi(\beta) \rangle \langle \psi(\beta)|$ for the state $|\psi(\beta)\rangle$ (Eq.\eqref{Eq:2d_toricCDA_Q}), one finds \cite{supplement}:
\begin{equation}
\label{Eq:tr_rhoA2_optimalCDA}
\begin{aligned}
&\tr(\rho_A^2) =
& \frac{ \sum_{\substack{x_\mathbf{e}, x'_\mathbf{e}}} \mathcal{Z}_{x_A, x_B}  \mathcal{Z}_{x_A', x_B'} e^{ -\Delta F_{AB}(x_\mathbf{e}, x'_\mathbf{e})/2} }{ \sum_{x_{\mathbf{e}}, x'_{\mathbf{e}}} \mathcal{Z}_{x_A, x_B} \mathcal{Z}_{x'_A, x'_B}}.
\end{aligned}
\end{equation}
Here $x_A (x_B)$ denotes all the edges belonging to the region $A(B)$, $\mathcal{Z}_{x_A, x_B}$ denotes the partition function of the 2d Ising model with the sign of Ising interactions determined by $x_A$ and $x_B$, and $\Delta F_{AB}(x_\mathbf{e}, x'_\mathbf{e}) = -\log [\mathcal{Z}_{x_A', x_B} \mathcal{Z}_{x_A, x_B'}/(\mathcal{Z}_{x_A, x_B}\mathcal{Z}_{x_A', x'_B})]$ is the free energy cost of swapping bonds between two copies of RBIM in region $A$. We provide a heuristic argument (Ref.\cite{supplement}) that the TEE jumps from $\log(2)$ to zero at $p_c = p_{\text{2d RBIM}}$. The main idea is that in the ferromagnetic phase of the RBIM, the free energy penalty of creating a single Ising vortex leads to a specific non-local constraint on the allowed configurations that contribute to the sum in Eq.\eqref{Eq:tr_rhoA2_optimalCDA}. The constraint is essentially that one needs to minimize the free energy cost $\Delta F_{AB}(x_\mathbf{e}, x'_\mathbf{e})$ for each fixed flux configuration  $\{f_p\}$ and $\{ f'_p\}$ corresponding to $\{x_e\}$ and $\{x'_e\}$ in Eq.\eqref{Eq:tr_rhoA2_optimalCDA}. One finds that there always exists a \textit{pair} of configurations that contribute equally to $\tr(\rho^2_A)$ while satisfying the aforementioned constraint. This results in a subleading contribution of $-\log(2)$ in the entanglement entropy, which we identify as TEE. In the paramagnetic phase, the aforementioned non-local constraint does not exist, and one therefore does not expect a non-zero TEE. Therefore, we arrive at the same conclusion as the one obtained using the anyon condensation operator.

Incidentally, one may also construct an alternative convex decomposition of the decohered mixed state $\rho$ that shows a phase transition at a certain (non-optimal) threshold $p_{\textrm{non-optimal}}$ which is related to 2d RBIM via a Kramers-Wannier duality \cite{kramers1941statistics}.  The main outcome is that $\tanh(\beta_\textrm{non-optimal}) = 1-2p_\textrm{non-optimal}$ satisfies $\tanh^2(\beta_\textrm{non-optimal}/2) = p_{\textrm{2d RBIM}}/ (1-p_{\textrm{2d RBIM}})$ which yields $p_\textrm{non-optimal} \approx 0.188$. See  Ref.\cite{supplement} for details.

Let us next consider the 3d toric code 	with $H_{\textrm{3d toric}} = - \sum_f (\prod_{e \in f} Z_e) - \sum_{v} (\prod_{e \in v } X_e)$ (see Fig.\ref{Fig:toric_xcube_main}(c)) subjected to phase-flip errors  (local non-trivial Kraus operators  $\sim Z_e$). Previous work has already identified error recovery transition at $p_c \approx 0.029$ with universality determined by the 3d random-plaquette gauge model (RPGM) along the Nishimori line \cite{wang2003confinement}. We first verify that the corresponding mixed-state density matrix is long-range entangled for $p < p_c$ by calculating its entanglement negativity. We find a non-zero, quantized topological negativity $\log(2)$ for $p < p_c$ and zero for $p > p_c$. The calculation is conceptually similar to that for the 2d toric code \cite{fan2023diagnostics}, see Ref.\cite{supplement} for details. Having confirmed the presence of long-range entanglement for $p < p_c$, we now ask whether the mixed-state at $p > p_c$ is separable?  To proceed, we follow a strategy similar to that for the 2d toric code, and rewrite the ground state of 3d toric code in terms of the ground state $\rho_{C,0}$ of a parent cluster state with Hamiltonian $H_{\textrm{3d cluster}}  = - \sum_e X_e \prod_{f \ni e} Z_f - \sum_f X_f \prod_{e \in f} Z_e  = \sum_e h_e + \sum_f h_f$. The corresponding ground state density matrix of the 3d toric code $\rho_{0}$ on a three-torus is
$\rho_{0} \propto \langle x_\mathbf{f} = 1| \rho_{C,0} |x_\mathbf{f} = 1\rangle$, which is an eigenstate of the non-contractible `t Hooft membrane operators $T_{xy}, T_{yz},T_{zx}$ along the three planes with eigenvalue +1 ($T_{xy} = \prod_{e \parallel z} X_e$ where the product is taken over all edges parallel to the $z$ axis in any $xy$ plane. $T_{yz}$ and $T_{zx}$ are defined analogously). 
Following essentially the same steps as in 2d toric code, one obtains the decohered density matrix $\rho =  \sum_{g_x} |\psi_{g_x}(\beta)\rangle \langle \psi_{g_x}(\beta)|$ with $|\psi_{g_x}\rangle = g_x |\psi(\beta)\rangle$ and $g_x$ a product of single-site Pauli-$X$s forming closed membranes. Therefore, we again only need to analyze whether $|\psi^{}\rangle \equiv \rho^{1/2}|z_\mathbf{e} = 1\rangle$ is SRE or LRE. Again, one may rewrite $|\psi(\beta)\rangle \propto \sum_{x_\mathbf{e}} [\mathcal{Z}_{\text{3d  gauge}, x_\mathbf{e}} (\beta)]^{1/2} |x_\mathbf{e}\rangle$ where  $\mathcal{Z}_{\text{3d  gauge},  x_\mathbf{e}} = \sum_{z_\mathbf{f}} e^{\beta \sum_e x_e \prod_{f \ni e} z_f}$ is now the partition function of a classical 3d Ising gauge theory with the sign of each plaquette term determined by $\{ x_e\}$. To probe the topological transition in $|\psi\rangle$ as function of $\beta$, we now consider the Wilson loop operator $W_{\ell} = \prod_{e \in \ell} Z_e$, where $\mathcal{\ell}$ denotes a homologically nontrivial cycle on the original lattice, say, along $z$ axis (so that it pierces and anti-commutes with $T_{xy}$). One finds $\langle  W_{\ell}  \rangle = \langle e^{-\Delta F_{\ell}/2} \rangle \geq e^{-\langle \Delta F_{\ell} \rangle /2}$, where $\Delta F_{\ell}$ now denotes the free energy cost of inserting a domain wall along the non-contractible loop for the 3d RPGM along the Nishimori line. Since $\langle  \Delta F_{\ell}  \rangle$ diverges as the length of $|\mathcal{C}| \sim L$ (= system-size) in the Higgs (ordered) phase, while converges to a constant in the confinement (disordered) phase \cite{wang2003confinement}, one finds $\langle W_{\ell} \rangle $ saturates to a non-zero constant when $p > p_{\text{3d RPGM}} \approx 0.029$ \cite{wang2003confinement}, while it vanishes for $p < p_c$. Therefore, $|\psi\rangle$, and correspondingly the decohered state $\rho$, is SRE when $p > p_{\text{3d RPGM}}$ and LRE for $p < p_{\text{3d RPGM}}$. One can similarly  study the 3d toric code with bit-flip errors. In this case, one finds that the separability transition is dictated by the transition out of the ferromagnetic phase in the 3d RBIM along the Nishimori line, which matches the optimal error-recovery threshold, $p_c \approx 0.233$ \cite{ozeki1998multicritical, kubica2018three}. See  Ref.\cite{supplement} for details.

Finally, let us briefly consider the 3d X-cube model  (Ref.\cite{vijay2016fracton}), where the Hilbert space consists of qubits residing on the edges(e) of a cubic lattice, and the Hamiltonian is $H_{\textrm{X-cube}} = - \sum_c \prod_{e \in c} Z_e - \sum_{v } (\prod_{e \in v_x } X_e + \prod_{e \in v_y } X_e +\prod_{e \in v_z } X_e )
= - \sum_c  A_c  - \sum_v (B_{v_x} + B_{v_y} + B_{v_z} )$ where $e \in v_\gamma,\ \gamma = x,y,z$ denotes all the edges emanating from the vertex $v$ that are normal to the $\gamma$-direction (see Fig.\ref{Fig:toric_xcube_main}(d)). Previous work has already established that under local decoherence, this system undergoes an error-recovery transition  at $p_c \approx 0.152$ with universality determined by 3d plaquette Ising model \cite{song2022optimal}. We now show that for $p > p_c$, the density matrix can be written as a convex sum of SRE pure states. We exploit the observation in Ref.\cite{verresen2021efficiently} that
the ground state density matrix $\rho_{0}$ of the X-cube model can be written as  $\rho_{0} \propto \langle x_\mathbf{c} = 1| \rho_{C,0} |x_\mathbf{c} = 1\rangle$ where $\rho_{C,0} = \prod_c(I-h_c) \prod_e (I-h_e)$ ($h_c = -X_c \prod_{e \in c} Z_e$ and $h_e = -X_e \prod_{c \ni e} Z_c$ denotes the ground state density matrix of the parent cluster state, and $|x_\mathbf{c} = 1\rangle = \otimes_c |x_c = 1\rangle$ is the product state in the Pauli-$X$ basis. The qubits in the parent cluster state live at the edges and the centers of the cubes so that $h_c$ involves 13-qubit interactions, and $h_e$ involves 5-qubit interactions (Fig.\ref{Fig:toric_xcube_main}(d)). The density matrix after subjecting $\rho_0$ to the phase-flip channel (Kraus operators $\sim Z_e$) can be written as $\rho \propto \sum_{x_\mathbf{e}} \mathcal{Z}_{\text{3d plaquette}, x_\mathbf{e}} |\Omega_{x_\mathbf{e}}\rangle \langle \Omega_{x_\mathbf{e}}| $, where $|\Omega_{x_\mathbf{e}}\rangle \propto \prod_e (I +  \prod_{e \in c} Z_e) |x_\mathbf{e}\rangle$ and $\mathcal{Z}_{\text{3d plaquette}, x_\mathbf{e}} = \sum_{z_\mathbf{c}} e^{\beta \sum_e x_e \prod_{c \ni e} z_c}$ is the partition function of the  3d plaquette Ising model \cite{savvidy1994geometrical} with the sign of interaction on each plaquette determined by $\{ x_e\}$.
One again only needs to analyze the state $|\psi\rangle = \rho^{1/2}|z_\mathbf{e} = 1 \rangle $ to study the separability transition for $\rho$. Now there exist exponentially many topological sectors \cite{vijay2016fracton}, and in the non-decohered ground state $\rho_0$, the membrane operators defined as $\prod_{e \parallel \hat{a}} X_e$ with $a = x,y,z$ for \textit{any} plane have expectation value one. To detect the presence/absence of topological order in $|\psi\rangle$, one therefore considers non-contractible Wilson loop operators $W_{\ell} = \prod_{e \in \ell} Z_e$ that anti-commute with the membrane operators orthogonal to $\ell$. 
The expectation value of any such Wilson loop takes a form similar to Eq.\eqref{Eq:noncont_wilson2d} where the partition function $\mathcal{Z}_{x_\mathbf{e}} = \mathcal{Z}_{\text{3d plaquette}, x_\mathbf{e}}$ and one is again along the Nishimori line. This again indicates that the pure state  $|\psi\rangle$  undergoes a transition at the error threshold $p_c = p_{\text{3d plaquette} } \approx 0.152$ \cite{song2022optimal}.

Finally, an alternative, heuristic approach to any of the phase transitions discussed above is via considering a more general class of wavefunctions $|\psi^{(\alpha)}\rangle \propto \rho^{\alpha/2}|z_\mathbf{e} = 1\rangle$, which capture the separability transition for the density matrix $\rho^{(\alpha)} \equiv \rho^{\alpha}/\tr(\rho^\alpha)$. For example, it is known that when $\rho$ is the decohered 2d toric code state, $|\psi^{(2)}\rangle$ undergoes a phase transition from being topologically ordered to an SRE state at a threshold  $p_c^{(2)}$ that is related to the critical temperature of the 2d \textit{translationally invariant} classical Ising model \cite{castelnovo2008quantum}, and correspondingly, we find that the topological negativity of $\rho^{(2)}$ undergoes a transition from $\log 2$ to $0$ at $p_c^{(2)}$ (see  Ref. \cite{supplement}). This transition can also be located by the wavefunction overlap $\log(\langle \psi^{(2)} (\beta)| \psi^{(2)} (\beta)\rangle)$, which is proportional to the free energy of the classical 2d Ising model. This motivates a generalization of this overlap to general $\alpha$ for any of the models considered above by defining $F_\alpha (\beta)= 
\frac{1}{1-\alpha} \log(\langle \psi^{(\alpha)} (\beta)| \psi^{(\alpha)} (\beta)\rangle)$.  Taking the limit $\alpha \to 1$, which corresponds to the wavefunction of our main interest (Eq.\eqref{Eq:2d_toricCDA_Q} for the 2d toric code, and analogous states for the other two models), one finds that $F_1 (\beta)$ precisely corresponds to the free energy of the corresponding statistical mechanics model along the Nishimori line, which indeed shows a singularity at the optimal error-recovery threshold $p_c$.


%

%

%

%

To summarize, we showed that decoherence-induced separability transitions in several topological states coincide with the optimal threshold for QEC \cite{dennis2002,fan2023diagnostics,bao2023mixed,lee2023quantum,song2022optimal}.  Therefore in these models, the inability to correct logical errors implies an ability to prepare the mixed state using an ensemble of short-depth unitary circuits, which is our main result. The parent cluster Hamiltonian approach we discuss also provides an alternative method to find the relevant statistical mechanics models. The convex decomposition discussed captures the universal aspects of the phase diagram, as well as the threshold correctly, and it is optimal in this sense. It will be interesting to consider generalization of our approach to non-CSS and/or non-abelian topological states. 
It might also be interesting to explore connections between decohered mixed states and perturbed \textit{pure} topological states since they are both connected to finite temperature classical phase transitions \cite{williamson2021stability}. Finally, we remark that using standard duality arguments \cite{wegner1971duality, lee2023quantum}, the separability transitions discussed here may also be reformulated as transitions for quantum magnets with a global $\mathbb{Z}_2$ symmetry. For example, our result implies that the ground state of the 2d quantum Ising model with Hamiltonian $H = - \sum_{\langle i,j\rangle} \tau^z_i \tau^z_j - h \sum_i \tau^x_i$ will undergo a separability transition in the paramagnetic phase when subjected to decoherence with Kraus operators $K_{ij} \propto \tau^z_i \tau^z_j$ at strength $p$. Above a certain $p_c$ (which,  at $h = 0$, equals $p_{\text{2d RBIM}} \approx 0.109$), one will enter a new phase where the density matrix is expressible  as a convex sum of states where each of them spontaneously breaks the $\mathbb{Z}_2$ symmetry, while such a representation is not possible below $p_c$ (recall that under Wegner duality, SRE states map to GHZ states). We leave a detailed exploration of this dual description for a future study.	

\begin{acknowledgments}
	The authors thank Dan Arovas,  Tim Hsieh and John McGreevy for helpful discussions, and Yimu Bao for helpful comments on the manuscript. TG is supported by the National Science Foundation under Grant No. DMR-1752417. This research was supported in part by the National Science Foundation under Grant No. NSF PHY-1748958.
\end{acknowledgments}

\bibliographystyle{apsrev4-2}

\newpage

\onecolumngrid

\appendix

\section{Entanglement entropy of the proposed optimal pure state for 2d toric code}

\label{sec:ES_optimalCDA} 

In this section, we compute the 2nd Renyi entanglement entropy of the (normalized) state in Eq.(2) of the main text, i.e., $|\tilde{\psi}\rangle = {\sum_{x_\mathbf{e}} \mathcal{\sqrt{Z}}_{x_\mathbf{e}}|x_\mathbf{e}\rangle}/{\sqrt{\sum_{x_\mathbf{e}}\mathcal{Z}_{x_\mathbf{e}}}}$ with $\mathcal{Z}_{x_\mathbf{e}} =  \sum_{z_\mathbf{v}} e^{\beta \sum_{e}  x_{e} \prod_{v \in e } z_v} $, the partition funcion of the 2d Ising model with the Ising interactions determined by $\{x_\mathbf{e}\}$.
We will later provide a heuristic argument that as one increases the error rate $p$ (which corresponds to increasing the temperature $\beta^{-1}$ in the Ising model via the relation $\tanh(\beta) = 1-2p$), the topological Renyi entropy undergoes a transition from $\log 2$ to $0$ at $p_c = p_{\text{2d RBIM}}$. We will only consider an entanglement bipartition such that the entangling boundary is contractible. Since the topological Renyi entropy for a contractible region does not depend on the topological sector $\mathbf{L}$ in which the wavefunction resides \cite{levin2006detecting,Kitaev06_1, dong2008topological,zhang2012quasiparticle}, below we will ignore the dependence of the partition function on $\mathbf{L}$ (alternatively, one may assume open boundary conditions so that there is a unique ground state).

The calculation of the 2nd Renyi entanglement entropy for a bipartition of the total system $A \cup B$ can be divided into two steps.
 First, we compute the reduced density matrix $\rho_A = \tr_B (|\tilde{\psi}\rangle \langle \tilde{\psi}|)$. Next, we compute $\tr\left(\rho^2_A\right)$ to get the 2nd Renyi  entropy $S_2 = -\log [\tr(\rho_A^2)]$.
Denoting $x_A (x_B)$ as the set of the links belonging to region $A(B)$, the pure state density matrix can be written as 
\begin{equation}
|\tilde{\psi}\rangle \langle \tilde{\psi}| =  \frac{ \sum_{x_\mathbf{e}, x'_\mathbf{e}} \sqrt{\mathcal{Z}_{x_A, x_B}}  \sqrt{\mathcal{Z}_{x_A', x_B'}} |x_A, x_B\rangle \langle x_A', x_B'|}{ (\sum_{x_e} \mathcal{Z}_{x_\mathbf{e}}) }. \nonumber
\end{equation}

\begin{figure}[t]
	\centering
	\includegraphics[width=0.5\linewidth]{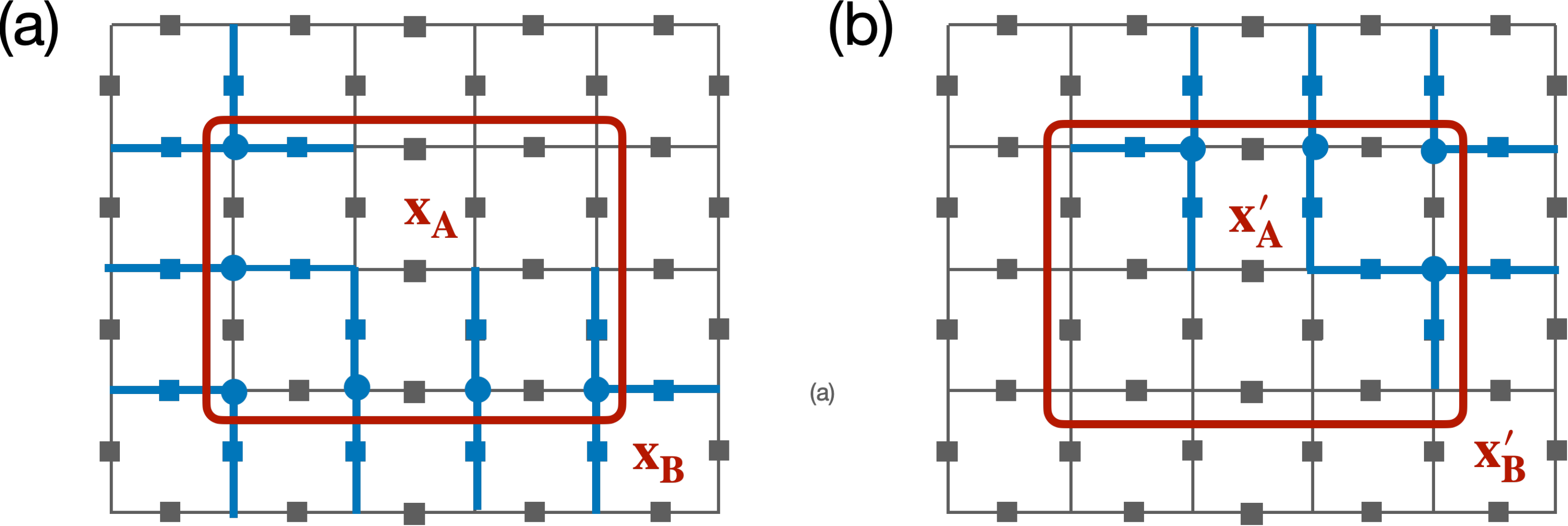}
	\caption{Given the boundary transformation $l_\mathbf{v}$ on $x_\mathbf{v}$ in (a), one can choose the boundary transformation $l'_\mathbf{v}$ on $x'_\mathbf{e}$ in (b) to be either $l'_\mathbf{v} = l_{\mathbf{v}}$(not shown) or $  l'_\mathbf{v} = \partial A +  l_\mathbf{v}\ (\text{mod}\ 2)$ to make the boundary fluxes $f_\mathbf{p}{(x'_A, x_B)}$ and $f_\mathbf{p}{(x_A, x'_B)}$  remain invariant.
		Recall that the  transformation is defined on vertices through $x_{e} \rightarrow x_{e} \prod_{v \in e} \sigma_v, \sigma_v = \pm 1$. Here we use blue circles to denote the vertices $\sigma_v = -1$ and blue squares to denote the edges where $ \prod_{v\in e} \sigma_v = -1$. } 
	\label{Fig:optimal_CDA_ES}
\end{figure}

Here $x_A (x_B)$ denotes all the edges belonging to the region $A(B)$ and $\mathcal{Z}_{x_A, x_B}$ denotes the partition of the 2d Ising model with the sign of Ising interactions determined by $x_A$ and $x_B$. Note that any given bond either belongs only to $A$ or only to $B$, see Fig.\ref{Fig:optimal_CDA_ES}.
One may then compute $\rho_A = \tr_B (|\tilde{\psi}\rangle \langle \tilde{\psi}|)$ to obtain
\begin{equation}
\begin{aligned}
\rho_A & = \frac{ \sum_{x_\mathbf{e}, x'_\mathbf{e}} \sqrt{\mathcal{Z}_{x_A, x_B}}  \sqrt{\mathcal{Z}_{x_A', x_B}} |x_A\rangle \langle x_A'|}{ (\sum_{x_e} \mathcal{Z}_{x_\mathbf{e}}) }.
\end{aligned}
\end{equation}
Therefore, 
\begin{align}
	\label{Eq:app_trrhoA2optimal}
	\begin{aligned}
		 \tr(\rho_A^2) & = \frac{ \sum_{\substack{x_\mathbf{e}, x'_\mathbf{e}}} \sqrt{\mathcal{Z}_{x_A, x_B} \mathcal{Z}_{x_A', x_B} \mathcal{Z}_{x_A, x_B'} \mathcal{Z}_{x_A', x_B'}} }{ \sum_{x_{\mathbf{e}}, x'_{\mathbf{e}}} \mathcal{Z}_{x_A, x_B} \mathcal{Z}_{x'_A, x'_B}} = \frac{ \sum_{\substack{x_\mathbf{e}, x'_\mathbf{e}}} \mathcal{Z}_{x_A, x_B}  \mathcal{Z}_{x_A', x_B'} e^{ -\Delta F_{AB}(x_\mathbf{e}, x'_\mathbf{e})/2} }{ \sum_{x_{\mathbf{e}}, x'_{\mathbf{e}}} \mathcal{Z}_{x_A, x_B} \mathcal{Z}_{x'_A, x'_B}} \\
	\end{aligned}
\end{align}
where 
\begin{equation}
\Delta F_{AB}(x_\mathbf{e}, x'_\mathbf{e}) = -\log (\frac{\mathcal{Z}_{x_A', x_B} \mathcal{Z}_{x_A, x_B'}}{\mathcal{Z}_{x_A', x'_B}\mathcal{Z}_{x_A, x_B}})
\end{equation}
is the free energy cost of swapping the bonds of two copies of RBIM (whose respective probabilities are given by $\mathcal{Z}_{x_A, x_B}$ and $\mathcal{Z}_{x_A', x'_B}$) in region $A$. 
Therefore, one finds that the second Renyi entropy $S_2 = -\log[\tr(\rho_A^2)]$ is related to the averaged free energy cost  along the Nishimori line of swapping the bonds of two copies of RBIM in region $A$.
The 2d RBIM along the Nishimori line exhibits two phases: a ferromagnetic phase when $p< p_{\text{2d RBIM}}$ and a paramagnetic phase when $p> p_{\text{2d RBIM}}$.
We now argue that the topological Renyi entropy is $\log 2$ ($0$) in the ferromagnetic (paramagnetic) phase. 

To begin with, let's first consider the extreme limit $\beta = \infty$.
Since $|\psi(\beta= \infty) \rangle$ is just the toric code ground state, we already know that the TEE is $\log 2$.
However, it is illuminating to see how this subleading correction emerges from Eq.\eqref{Eq:app_trrhoA2optimal} in this limit. We first note that the probability for a given configuration $(x_\mathbf{e}, x_\mathbf{e}')$ to occur is completely determined by  $\mathcal{Z}_{x_A, x_B} \mathcal{Z}_{x'_A, x'_B} $, which is merely a function of fluxes $f_{\mathbf{p}}(x_\mathbf{e})$ and  $f_{\mathbf{p}}(x_\mathbf{e}')$, where we denote $f_{\mathbf{p}}(x_\mathbf{e}) = \{f_p (x_\mathbf{e})  \}$ as the flux configuration determined by $x_\mathbf{e}$ (recall that $f_p = \prod_{e \in p}x_e$). As noted in the first paragraph of this appendix, if the system is on a torus, then although  $\mathcal{Z}$ technically also depends on the logical sector $\mathbf{L}$, such dependence is not important for our discussion here, as we are working with a region $A$ whose boundary is contractible.
When $\beta = \infty$, the partition function belonging to the vacuum sector (i.e., $ f_p = 1, \forall p$) dominates. 
Therefore, the dominant terms in the denominator of Eq.\eqref{Eq:app_trrhoA2optimal} are the configurations $(x_\mathbf{e}, x_\mathbf{e}')$ that ensure $f_p(x_\mathbf{e}) = f_p (x_\mathbf{e}') = 1,\ \forall p$.
On the other hand, the dominant terms in the numerator of Eq.\eqref{Eq:app_trrhoA2optimal}  should satisfy the extra condition that $\Delta F_{AB}(x_\mathbf{e}, x_\mathbf{e}') = 0$.
This forces the swapped configurations $(x_A', x_B)$  and $(x_A, x_B')$ to also lie in the vacuum sector, i.e., $f_p(x'_A, x_B) = f_p(x_A, x_B') =1, \forall p$, where we denote $f_p(x'_A, x_B)  $ as the flux determined by the bonds $(x'_A, x_B)$.
Therefore, Eq.\eqref{Eq:app_trrhoA2optimal} in the limit $\beta = \infty$ simply counts the degrees of freedom lost due to the additional constraint $\Delta F_{AB}(x_\mathbf{e}, x_\mathbf{e}') = 0$ in the vacuum sectors of $x_\mathbf{e}$ and $x_\mathbf{e}'$.

Based on above considerations, to evaluate Eq.\eqref{Eq:app_trrhoA2optimal} for the case $\beta = \infty$, we recall that different configurations $x_\mathbf{e}$ in the vacuum sector can be obtained by applying the transformation (defined on the vertices) $x_{e} \rightarrow x_{e}  \prod_{v \in e} \sigma_v, \sigma_v = \pm 1$ to the disorder-free sector, i.e., $x_e = 1, \forall e$.
Therefore, without the constraint $\Delta F_{AB}(x_\mathbf{e}, x_\mathbf{e}') = 0$, there are $2^{2|\partial A|}$ different transformations on the boundary vertices (i.e., the vertices living on the boundary between region $A$ and $B$) that make $x_\mathbf{e}$ and $ x_\mathbf{e}'$ stay in the vacuum sectors, where $|\partial A|$ denotes the number of boundary vertices (for example, $|\partial A| = 10$ in Fig.\ref{Fig:optimal_CDA_ES}).
On the other hand, after enforcing the constraint $\Delta F_{AB} (x_\mathbf{e}, x_\mathbf{e}') = 0$, the transformation on the boundary vertices for $x_\mathbf{e}$ and $x_\mathbf{e}'$ are no longer independent, and can be counted as follows (note that $\Delta F_{AB} (x_\mathbf{e}, x_\mathbf{e}') $ will only be modified by the transformations on the boundary, so the transformations in the bulk for $x_\mathbf{e}$ and $x_\mathbf{e}'$ are not subjected to any constraint).
We first note that the disorder-free configuration for both $x_\mathbf{e}$ and $x_\mathbf{e}'$ of course satisfy $\Delta F_{AB}(x_\mathbf{e}, x_\mathbf{e}') = 0$.
Starting from the disorder-free configuration, there are $2^{|\partial A|}$ transformations (including the trivial transformation) that can be applied to $x_\mathbf{e}$ so that $f_p(x_\mathbf{e}) = 1, \forall p$.
Crucially, given a specific boundary transformation $l_\mathbf{v}$ applied to $x_\mathbf{e}$, there are \textit{two}  boundary transformations $l'_\mathbf{v}$ for $x'_\mathbf{e}$ that make the swapped configurations $(x_A', x_B)$ and $(x_A, x_B')$ remain in the vacuum sector: either $l'_\mathbf{v} = l_{\mathbf{v}}$ or $l'_\mathbf{v}  = \partial A  + l_\mathbf{v}\ (\text{mod}\ 2)$.
We provide one such example in Fig.\ref{Fig:optimal_CDA_ES}.
As a consequence, there are $2^{|\partial A|} \times 2$ transformations on the boundary that satisfy $\Delta F_{AB}(x_\mathbf{e}, x_\mathbf{e}') = 0$.
It follows that 
\begin{equation}
\begin{aligned}
S_2 &  = -\log(\frac{2^{|\partial A| +1} }{ 2^{2|\partial A|} }) = -\log(\frac{1 }{ 2^{|\partial A| - 1} })= (|\partial A|  - 1) \log 2,
\end{aligned}
\end{equation}
which is precisely the entanglement entropy of the 2d toric code ground state, as expected.

Given the aforementioned understanding in the limit $\beta = \infty$, let's now consider the whole ferromagnetic phase $\beta > \beta_c$, where $\beta_c$ is the critical temperature of the RBIM along the Nishimori line. Since now $\beta$ is no longer infinite, all  flux configurations $f_{\mathbf{p}}{(x_\mathbf{e})}$ and $f_{\mathbf{p}}{(x_\mathbf{e}')}$ contribute in Eq.\eqref{Eq:app_trrhoA2optimal}.
However, crucially, the free energy cost of changing a single flux (i.e., $f_p \rightarrow -f_p $ for a given $p$) scales with the system size in the ferromagnetic phase \cite{dennis2002}.
%
%
Therefore, for a given $(f_\mathbf{p}, f'_\mathbf{p} )$, it is reasonable to expect that the numerator in Eq.\eqref{Eq:app_trrhoA2optimal} is dominated by the configurations that minimize $\Delta F_{AB} (x_\mathbf{e}, x_\mathbf{e}')$, where $(x_\mathbf{e}, x_\mathbf{e}') \in (f_\mathbf{p}, f'_\mathbf{p} )$.
Similar to the counting of the vacuum sector in the limit $\beta = \infty$, all such configurations can be obtained by applying transformations akin to Fig.\ref{Fig:optimal_CDA_ES} to a configuration $(x_{\mathbf{e},0}, x_{\mathbf{e},0}') \in (f_\mathbf{p}, f'_\mathbf{p} )$ that minimizes $\Delta F_{AB} $.
It follows that there are again $2^{|\partial A| +1}$ boundary transformations for each $(f_\mathbf{p}, f'_\mathbf{p} )$, leading to a $-\log 2$ subleading correction.
We note that since the minimized free energy $\Delta F_{AB}(x_\mathbf{e} \in f_\mathbf{p}, x_\mathbf{e}' \in f'_\mathbf{p})$ depends on the flux sector $(f_\mathbf{p}, f'_\mathbf{p})$, the leading boundary-law coefficient will be modified.
Therefore, we expect the second Renyi entropy $S_2 = \alpha |\partial A| - \log 2$ in the ferromagnetic phase, with $\alpha$ a non-universal constant.
On the other hand, when $\beta < \beta_c$, so that the system is in the paramagnetic phase, the free energy cost of changing a single flux is finite. 
Therefore, there is no non-local constraint on the configurations in Eq.\eqref{Eq:app_trrhoA2optimal}, and therefore, one expects a purely boundary-law scaling of the Renyi entropy without any subleading TEE.

We note that our argument for TEE is  reminiscent of the argument in Refs.\cite{castelnovo2008quantum,fan2023diagnostics,bao2023mixed} where non-zero TEE/topological negativity results from a non-local constraint on the entanglement boundary. However, the precise origin of the constraint is a bit different, as the calculation in Refs.\cite{castelnovo2008quantum,fan2023diagnostics,bao2023mixed} is in a dual picture where the topological (trivial) phase corresponds to a paramagnetic (ferromagnetic) phase.

\section{A non-optimal decomposition}
\label{sec:nonoptimal_CDA}
Here we provide details of the non-optimal decompositions for the 2d and 3d toric codes subjected to the phase-flip errors whose results are briefly mentioned in the main text. 

\subsubsection*{2d toric code}
\begin{figure*}
	\centering
	\includegraphics[width=0.9\linewidth]{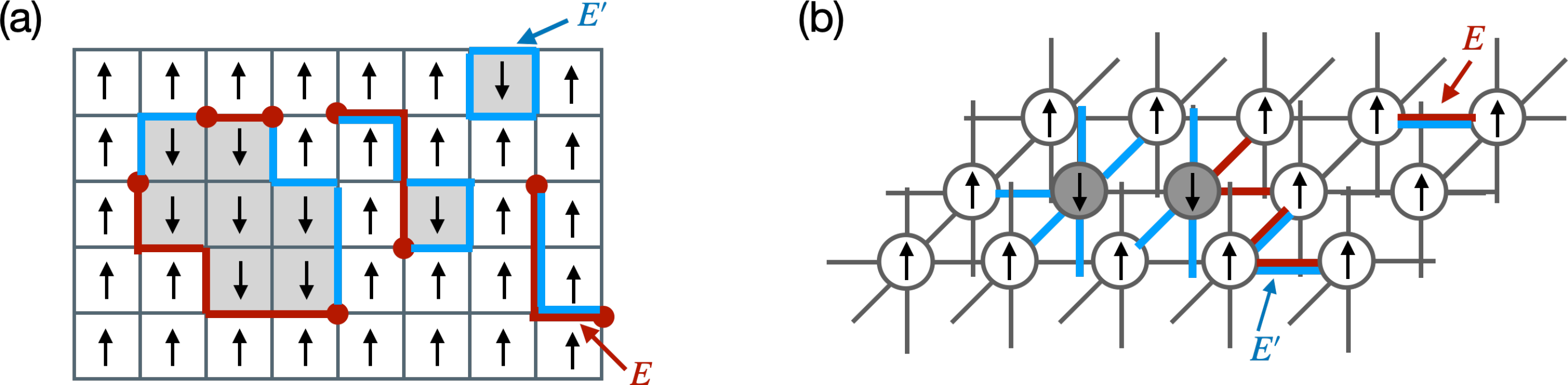}
	\caption{
		The correspondence between the components of the (non-optimal) convex decomposition for  toric codes, i.e., $[\tanh(\beta/2)]^{|E'|}\prod_{e \in E'} X_e |z_\mathbf{e}\rangle$ (see Eq.\eqref{Eq:toric_METS_RBIM}) and the domain wall configurations of the Ising model with the Ising interaction determined by $\{ z_{e} \}$ in (a) 2d and (b) 3d.
		Note that the classical spins reside on plaquettes for case (a) while they reside on vertices for case (b).
		The edges with $z_e = -1$ of the initial state $|z_\mathbf{e}\rangle $  label the chain $E$ of antiferromagnetic bonds (red).
		The edges where Pauli-$X$s are applied to correspond to the chain $E'$ of excited bonds (blue), which is determined by $E' = D+ E$ ($\textrm{mod}\ 2$) with $D$ labeling the domain walls (the boundary of the shaded regions).
		The squared norm of the amplitude, namely $[\tanh^2(\beta/2)]^{|E'|}$, corresponds to the probability for such a domain wall configuration.  This notation is directly borrowed from Ref.\cite{dennis2002}.
	} 
	\label{Fig:toric_both}
\end{figure*}
To obtain an alternative convex decomposition of the decohered toric code density matrix in Eq.(1) of the main text, we again return to the bigger Hilbert space of the 2d cluster state, and write the decohered state $\rho$  as $\rho \propto \Gamma \Gamma^\dagger $, where
\begin{equation}
	\label{Eq:G_non_optimal}
	\Gamma = \langle  x_\mathbf{v} = 1|
	e^{-\beta \sum_e h_{e}^{}/2}\prod_v (I -h_v^{})
\end{equation}
(recall $h_v = - X_v \prod_{e \ni v} Z_e$ and $h_e = - X_e \prod_{v \in e} Z_v$ for the 2d cluster Hamiltonian).
We note that unlike the operator $\rho^{1/2}$ that was used for the optimal decomposition in the main text, $\Gamma$ in Eq.\eqref{Eq:G_non_optimal} is not a square matrix, but instead is of size $2^{N_e } \times 2^{N_e + N_v}$, where $N_e \,(N_v)$ denotes the number of edges (vertices) of the square lattice on which the cluster state lives. 
To proceed, we insert the complete basis $\{ |m\rangle = |z_\mathbf{e}, x_\mathbf{v}\rangle \}$ between $\Gamma$ and $\Gamma^\dagger$, where $|z_\mathbf{e} \rangle= \otimes_e|z_{e}\rangle$ ($|x_\mathbf{v}\rangle = \otimes_v|x_{v}\rangle$) denotes the eigenvector of Pauli-$Z_e$ (Pauli-$X_v$) with eigenvalue $z_{e} \,(x_v) = \pm 1$.
The corresponding $|\psi_m\rangle$ for a gvien $m$ can then be written as $|\psi_m\rangle \propto \Gamma |z_\mathbf{e}, x_\mathbf{v}\rangle$.
Since $\prod_v (I - h_v^{})$ is the projector to the even sector of $X_v (\prod_{e \ni v} Z_e)$, $|\psi_m\rangle  $ is non-vanishing only when $x_v = \prod_{e \in v} z_e$. 
Therefore, $x_\mathbf{v}$ is completely determined by $z_\mathbf{e}$.
It is then more natural to label $|\psi_m\rangle$ using $z_\mathbf{e}$, and each $|\psi_{z_\mathbf{e}} \rangle$ can be written as
\begin{equation}
	\label{Eq:2d_nonoptimal}
	\begin{aligned}
		|\psi_{z_\mathbf{e}}\rangle \propto  \langle x_\mathbf{v} = 1| e^{-\beta \sum_e h_{e}^{} /2 } |z_\mathbf{e},  x_\mathbf{v} = \prod_{e \in \mathbf{v} } z_e \rangle. 
	\end{aligned}
\end{equation}
We emphasize that unlike the optimal decomposition mentioned in the main text (namely, $\rho^{1/2} |z_\mathbf{e}\rangle, \forall z_\mathbf{e}$), the states $|\psi_{z_\mathbf{e}}\rangle$ that enter the convex decomposition here are \textit{not} related to each other via a product of single-site unitaries. This implies that one should analyze all the $|\psi_{z_\mathbf{e}}\rangle$ to determine whether $\rho = \sum_{z_\mathbf{e}} p_{z_\mathbf{e}} |\psi_{z_\mathbf{e}}\rangle \langle \psi_{z_\mathbf{e}}|$ is a convex sum of SRE states.
Besides, since the number of  states $|\psi_{z_\mathbf{e}}\rangle$ grows exponentially as a function of the system size, the probability for each $|\psi_{z_\mathbf{e}}\rangle$ is exponentially small in general.
It follows that even if there exists a subset of $|\psi_{z_\mathbf{e}}\rangle$ that are not SRE, the decohered state $\rho$ may still be well approximated by an SRE mixed state, as long as the total probability of the LRE pure states is vanishingly small.
Therefore, the notion of $\rho$ being a SRE mixed state must take into account the probability for each $|\psi_{z_\mathbf{e}}\rangle$, and can only be made precise in a statistical sense. 
To make progress, we characterize the ensemble $\{ |\psi_{z_\mathbf{e}}\rangle\}$ by the following \textit{average} anyon condensation parameter:
\begin{equation}
	\label{Eq:average_anyon_condense}
	[\langle T_{\tilde{\mathcal{C}}} (\beta) \rangle^2] \equiv \sum_{z_\mathbf{e}} p_{z_\mathbf{e}}(\beta) \langle T_{\tilde{\mathcal{C}}}(\beta) \rangle_{z_\mathbf{e}}^2,
\end{equation}
where $T_{\tilde{\mathcal{C}}} = \prod_{e \in \tilde{\mathcal{C}}} Z_e$ with $\tilde{\mathcal{C}}$ an open curve on the dual lattice, and 
$\langle T_{\tilde{\mathcal{C}}} (\beta)\rangle_{z_\mathbf{e}} \equiv \langle \psi_{z_\mathbf{e}}|T_{\tilde{\mathcal{C}}}|\psi_{z_\mathbf{e}}\rangle/\langle \psi_{z_\mathbf{e}}| \psi_{z_\mathbf{e}}\rangle$.
%
Intuitively, $T_{\tilde{\mathcal{C}}}$ can be understood as an operator that creates a pair of anyons on the plaquettes $p$ and $p'$, the end points of the curve $\tilde{\mathcal{C}}$. 
Therefore, the expectation value of $T_{\tilde{\mathcal{C}}} $ vanishes for a topologically ordered (LRE) state as $|\tilde{\mathcal{C}}| \rightarrow \infty$ and is nonzero for a non-topologically ordered/anyon condensed (SRE) state.  We note that a similar order parameter has also been employed in Ref.\cite{bao2023mixed} to understand the decohered mixed state using C-J map/double state formalism.
Our main result is the following equality:
\begin{equation}
	\label{Eq:RC_ss}
	[\langle T_{\tilde{\mathcal{C}}} (\beta) \rangle^2]= [\langle s_p s_{p'}(K) \rangle]_{\text{2d RBIM}},\ K = -\log \tanh(\beta/2),
\end{equation}
where $[\langle s_p s_{p'}(K) \rangle]_{\text{2d RBIM}}$ denotes the disorder-averaged spin-spin correlation function of the 2d RBIM along the Nishimori line at inverse temperature $K$.
Crucially, due to the relation $e^{-K} = \tanh(\beta/2)$, the low $K$ limit (paramagnetic phase) now corresponds to the large $\beta$ (low error rate) limit, in contrast to the discussion of the optimal decomposition in the main text where the low error rate corresponds to the ferromagnetic phase. This is a consequence of the Kramers-Wannier duality. Therefore, $[\langle T_{\tilde{\mathcal{C}}} \rangle^2]$ vanishes in the low error rate limit, consistent with our expectation. To determine the critical error rate $p_{\textrm{non-optimal}}$ i.e. the error rate beyond which the decohered density matrix is (statistically) SRE, we use $\tanh(\beta_\textrm{non-optimal}) = 1-2p_\textrm{non-optimal}$ and $\tanh^2(\beta_\textrm{non-optimal}/2) = e^{-2K_{\textrm{2d RBIM}}} = p_{\textrm{2d RBIM}}/ (1-p_{\textrm{2d RBIM}})$ with $p_{\textrm{2d RBIM}} \approx 0.109$ to obtain $p_\textrm{non-optimal} \approx 0.188$.
As expected, $p_\textrm{non-optimal} > p_\textrm{optimal} = p_{\textrm{2d RBIM}}$.

The sketch of our calculations that verify Eq.\eqref{Eq:RC_ss} is as follows.
We will first compute the probabilty $p_{z_\mathbf{e}}$ for each $|\psi_{z_\mathbf{e}}\rangle$ to occur, which is simply proportional to the normalization of $|\psi_{z_\mathbf{e}}\rangle$.
We then evaluate the expectation value of anyon condensation operator $\langle T_{\tilde{\mathcal{C}}} \rangle_{z_\mathbf{e}}$.
It will then become transparent that Eq.\eqref{Eq:average_anyon_condense} is equivalent to the 2d RBIM along the Nishimori line.

 To compute the probability $p_{z_\mathbf{e}} \propto \langle \psi_{z_\mathbf{e}}|\psi_{z_\mathbf{e}}\rangle$, we first simplify the expression of $|\psi_{z_\mathbf{e}}\rangle$. It is illuminating to denote the edges where $z_e = -1$ as the string configuration $E$ so that $x_v = \prod_{e \in v} z_e = -1$ corresponds to the end points of $E$ (see Fig.\ref{Fig:toric_both}(a)).
Now, expand $e^{-\beta \sum_e h_{e} /2 } \propto \prod_e (I - \tanh(\beta/2) h_e)$ as a sum of several string operators generated by product of $h_e$s, which we denote as $E'$.
For $E'$ to survive after the projection $\langle x_\mathbf{v} = 1|$, its end points should coincide with the end points of $E$ ($x_v = -1$) so that $h_e$ flip those spins to $x_v = +1$.
This implies that $E' + E$ ($\textrm{mod}\ 2$) should form closed loop configurations $D$, and $|\psi_{z_\mathbf{e}}\rangle $ for a given ${z_\mathbf{e}}$ (which fully specifies $E$) can be written as
\begin{equation}
	\label{Eq:toric_METS_RBIM}
	\begin{aligned}
		|\psi_{z_\mathbf{e}}\rangle \propto \sum_{E' \textrm{s.t.} E' + E = D} [\tanh(\beta/2)]^{|E'|} \prod_{e \in E'}X_e |z_\mathbf{e}\rangle.
	\end{aligned}
\end{equation}
Since each $\prod_{e \in E'}X_e |z_\mathbf{e}\rangle$ is orthogonal to one another, the probability $p_{z_\mathbf{e}} \propto \langle \psi_{z_\mathbf{e}}|\psi_{z_\mathbf{e}}\rangle$  can be evaluated straightforwardly as 
\begin{equation}
	p_{z_\mathbf{e}} \propto  \sum_{E' \textrm{s.t.} E' + E = D} [\tanh^2(\beta/2)]^{|E'|},
\end{equation}
which is nothing but the partition function of 2d Ising model $\mathcal{Z}_{\textrm{2d Ising}, {z_\mathbf{e}}}(K)$ at inverse temperature $K = -\log \tanh(\beta/2)$ and the negative Ising interactions labeled by $E'$ (determined by ${z_\mathbf{e}}$).
Besides, there is a one-to-one mapping between the components of $|\psi_{z_\mathbf{e}}\rangle$, i.e., $[\tanh(\beta/2)]^{|E'|}\prod_{e \in E'} X_e |z_\mathbf{e}\rangle$, and the domain wall configurations of a statistical mechanics model. Specifically, the classical spins reside on the plaquettes of the square lattice, and the bonds where $z_e = -1$ in the initial state $|z_\mathbf{e}\rangle $  label the chain $E$ of the antiferromagnetic bonds.  The end points of $E$ labeled by $x_v = \prod_{e \in v} z_e = -1$ then correspond to the Ising vortices. The edges where Pauli-$X$s are applied correspond to the chain $E'$ of excited bonds, which is determined by $E' = D+ E $ ($\textrm{mod}\ 2$). The squared norm of the amplitude  $[\tanh^2(\beta/2)]^{|E'|} = e^{-2K|E'|}$ corresponds to the probability for such a domain wall configuration to appear.   We summarize this correspondence in Fig.\ref{Fig:toric_both}(a).

Now, we are ready to compute $\langle T_{\tilde{\mathcal{C}}}  \rangle_{z_\mathbf{e}}$ for a given $z_\mathbf{e}$.
Using $T_{\tilde{\mathcal{C}}}  (\prod_{e\in E'}X_e) = (-1)^{\#\textrm{cross}(E', \tilde{\mathcal{C}})} (\prod_{e \in E'}X_e) T_{\tilde{\mathcal{C}}} $ and $T_{\tilde{\mathcal{C}}} |z_\mathbf{e}\rangle = (-1)^{\#\textrm{cross}(E, \tilde{\mathcal{C}})}|z_\mathbf{e}\rangle$ with $\#\textrm{cross}(A,B)$ the number of crossings between $A$ and $B$, we get 
\begin{equation}
	\begin{aligned}
		T_{\tilde{\mathcal{C}}} |\psi_{z_\mathbf{e}}\rangle
		=  \sum_{E'} (-1)^{\#\textrm{cross}(D, \tilde{\mathcal{C}})}  e^{-K|E'|}\prod_{e \in E'} X_e |z_\mathbf{e}\rangle,
	\end{aligned}
\end{equation}
where we have used $D = E + E'$.
It  follows that
\begin{equation}
	\langle T_{\tilde{\mathcal{C}}}  (\beta)\rangle_{z_\mathbf{e}}  
	 = 
	\frac{\sum_{E'} (-1)^{\#\textrm{cross}(D, \tilde{\mathcal{C}})}  e^{-2K|E'|}}{\mathcal{Z}_{\textrm{2d Ising}, {z_\mathbf{e}}}(K)},
\end{equation}
which corresponds to the spin-spin correlation function $\langle s_{p} s_{p'} (K)\rangle$ of the 2d Ising model with the Ising interaction determined by $\{ z_e \}$, where $p$ and $p'$ denote the end points of $\tilde{\mathcal{C}}$.
It follows that we can characterize the separability transition for the mixed state $\rho = \sum_{z_\mathbf{e}} p_{z_\mathbf{e}}|\psi_{z_\mathbf{e}}\rangle \langle \psi_{z_\mathbf{e}}|$ using Eq.\eqref{Eq:average_anyon_condense}:
\begin{equation}
\begin{aligned}
	[\langle T_{\tilde{\mathcal{C}}} \rangle^2](\beta) & = \sum_{z_\mathbf{e}} p_{z_\mathbf{e}} \langle T_{\tilde{\mathcal{C}}}  \rangle^2_{z_\mathbf{e}}  =  \frac{\sum_{z_\mathbf{e}} \mathcal{Z}_{\text{2d Ising}, {z_\mathbf{e}}}(K) \langle s_p s_p'(K)\rangle}{\sum_{z_\mathbf{e}} \mathcal{Z}_{\text{2d Ising}, {z_\mathbf{e}}}(K)}.
\end{aligned}
\end{equation}
Therefore, as claimed in Eq.\eqref{Eq:RC_ss}, $[\langle T_{\tilde{\mathcal{C}}}  \rangle^2](\beta)$  precisely corresponds to the disorder-averaged spin-spin correlation function of the RBIM along the Nishimori line with inverse temperature $K = -\log \tanh(\beta/2)$.

\subsubsection*{3d toric code}
Here we derive a non-optimal decomposition for the density matrix of the 3d toric code subjected to phase-flip errors. Similar to the 2d toric code, we write $\rho \propto \Gamma \Gamma^\dagger$ with 
\begin{equation}
	\Gamma = \langle x_\mathbf{f} = 1|
	e^{-\beta \sum_e h_{e}/2} \prod_f (I - h_f)
\end{equation}
(recall $h_e = -X_e \prod_{f \ni e} Z_f$  and $h_f = -X_f \prod_{e \in f} Z_e $ for the 3d cluster Hamiltonian).
This equation motivates the decomposition $\rho = \sum_m |\psi_m\rangle \langle \psi_m|$, where $|\psi_m\rangle \propto \Gamma|z_\mathbf{e}, x_\mathbf{f}\rangle$ and $|z_\mathbf{e} \rangle= \otimes_e|z_{e}\rangle$ ($|x_\mathbf{f}\rangle = \otimes_f|x_{f}\rangle$) denotes the eigenvector of Pauli-$Z_e$ (Pauli-$X_f$) with eigenvalue $z_{e} (x_f) = \pm 1$. 
Since $\prod_f (I - h_f)$ is the projector to the even sector of $X_f (\prod_{e \in f} Z_e)$, $|\psi_m\rangle$ is non-vanishing only when $x_f = \prod_{e \in f} z_e$. 
Therefore, $x_\mathbf{f}$ is completely determined by $z_\mathbf{e}$, and it is more natural to label $|\psi_m\rangle$ using $z_\mathbf{e}$, which can be written as
\begin{equation}
	\label{Eq:nonoptimal_3dtoric}
	\begin{aligned}
		|\psi_{z_\mathbf{e}}\rangle \sim  \langle x_\mathbf{f} = 1| e^{-\beta \sum_e h_{e}^{} /2 } |z_\mathbf{e},  x_\mathbf{f} = \prod_{e \in \mathbf{f} } z_e \rangle. 
	\end{aligned}
\end{equation}
To statistically characterize the separability condition for $\rho$, one can use the anyon condensation operator which in this case is a `Wilson string operator' $W_{\mathcal{C}}$ where the open curve $\mathcal{C}$ lives on the original lattice (note that unlike a proper $\mathbb{Z}_2$  theory, $W_{\mathcal{C}}$  here does not vanish identically since the Gauss law is being enforced  only energetically, and not exactly).
Specifically, our main result is:
\begin{equation}
	\label{Eq:3d_WC}
	[\langle W_{\mathcal{C}}(\beta) \rangle^2]= [\langle s_p s_{p'}(K) \rangle]_{\text{3d RBIM}},\ K = -\log \tanh(\beta/2),
\end{equation}
where $[\langle s_p s_{p'}(K) \rangle]_{\text{3d RBIM}}$ denotes the disorder-averaged spin-spin correlation function of the 3d RBIM along the Nishimori line with inverse temperature $K$.
We note that the 3d Ising model is dual to the 3d Ising plaquette gauge model, consistent with our expectation that the non-optimal state is related to the optimal one via a Kramers-Wannier duality.
To determine the bound on the error rate $p_\textrm{non-optimal}$ for separability of $\rho$ resulting from this calculation, we use $\tanh(\beta_\textrm{non-optimal}) = 1-2p_{\textrm{non-optimal}}$ and $\tanh^2(\beta_\textrm{non-optimal}/2) = e^{-2K_{\textrm{3d RBIM}}} = p_{\textrm{3d RBIM}}/ (1-p_{\textrm{3d RBIM}})$ with $p_{\textrm{3d RBIM}} \approx 0.233$ \cite{ozeki1998multicritical} to obtain $p_\textrm{non-optimal} \approx 0.077$. Therefore, the outcome of this analysis is that the decohered density matrix is SRE for $p > p_\textrm{non-optimal}$. As expected, this non-optimal bound is higher than the optimal threshold $p_\text{3d RPGM} \approx 0.029$ for separability argued in the main text.

To verify Eq.\eqref{Eq:3d_WC}, we will follow the same derivation as for the 2d toric code by first 
computing the probabilty $p_{z_\mathbf{e}}$ and then evaluating $\langle W_{{\mathcal{C}}} \rangle_{z_\mathbf{e}}$.
Similar to the 2d toric code, we denote the edges with $z_e = -1$ as $E$. Expanding $e^{-\beta \sum_e h_{e}^{} /2 } |z_\mathbf{e},  x_\mathbf{f} = \prod_{e \in \mathbf{f} } z_e \rangle \propto \prod_f (I - \tanh(\beta/2) h_e^{})|z_\mathbf{e},  x_\mathbf{f} = \prod_{e \in \mathbf{f} } z_e \rangle$  yields a superposition of terms in the $Z$-basis, and we denote the corresponding configurations as $E'$. To see what subset of $E'$ survives after the projection $\langle x_f = 1|$, we note that when one applies $h_{e}^{}$, all the spins $x_f$ with $f \ni e$ will be flipped. Therefore, $h_e$ acts non-trivially only on faces $f$, where $x_f = -1$ with  $e \in f$.  The condition $x_f = -1$ is satisfied precisely on faces that are bounded by edges $e$ with an odd number of $z_e = -1$ (since $x_f = \prod_{e \in f} z_e$). This implies that $E+E'$ should form closed surfaces (denoted as $D$). Therefore, $|\psi_{z_\mathbf{e}}\rangle$ for a given $z_\mathbf{e}$ (which fully specifies $E$) has the same form as Eq.\eqref{Eq:toric_METS_RBIM}, though here $D$ represents closed surfaces instead of closed loops. 
 Since each $\prod_{e \in E'} X_e |z_\mathbf{e}\rangle$ is orthogonal to one another, $p_{z_\mathbf{e}} \propto \langle \psi_{z_\mathbf{e}}|\psi_{z_\mathbf{e}}\rangle$ can be computed as $\sum_{E' \textrm{s.t.} E' + E = D} [\tanh^2(\beta/2)]^{|E'|}$,
which is nothing but the partition function of 3d Ising model $\mathcal{Z}_\textrm{3d Ising}(K)$ with inverse temperature $K = -\log \tanh(\beta/2)$ and the negative Ising interactions labeled by $E'$ (determined by ${z_\mathbf{e}}$).
Therefore, similar to the 2d toric code, each component of the non-optimal $|\psi_{z_\mathbf{e}}\rangle$ has a one-to-one correspondence with the domain-wall configuration of the 3d Ising model, which we summarize in Fig.\ref{Fig:toric_both}(b).
A derivation similar to that for the 2d toric code then shows that the average Wilson string operator (Eq.\eqref{Eq:3d_WC})  is equivalent to the disorder-averaged correlation function of the 3d RBIM along the Nishimori line with inverse temperature $K = -\log \tanh(\beta/2)$.

\section{Entanglement negativity of 3d toric code under phase-flip errors}
The goal of this appendix is to show that the topological entanglement negativity of 3d toric code under phase-flip errors is $\log(2)$ for $p < p_c = p_{\text{3d RPGM}} \approx 0.029$ and vanishes above $p_c$.
The overall strategy is similar to the one in Ref.\cite{fan2023diagnostics}:  we first map the $2n$-th Renyi entanglement negativty 
\begin{equation}
	\label{Eq:renyi_negativity_def}
	\mathcal{N}_A^{(2n)} = \frac{1}{2-2n} \log \frac{\tr[(\rho^{T_A})^{2n}]}{\tr(\rho^{2n}) },
\end{equation}
where $T_A$ denotes the partial transpose on region $A$, to the excess free energy of the corresponding statistical mechanical model with a certian constraint. 
We then argue that such a quantity, aside from the area-law contribution, has a subleading $\log 2$ correction when $p < p_c^{(2n)}$ with $ p_c^{(2n)}$ the critical error rate of the  corresponding statistical mechanical model. Assuming the validity of the replica limit  $2n \rightarrow 1$, it then follows
that the topological entanglement negativity  vanishes above $p >p_c^{(1)} =  p_{\text{3d RPGM}}$.

Similar to the 2d toric code, one can express the 3d toric code wave function as an equal weight superposition of closed surfaces:
\begin{equation}
	|\Psi_0\rangle = \prod_v (I + \prod_{e \in v} X_e)|0\rangle = \sum_{g} g_x |0\rangle = \sum_{g}  |g\rangle.
\end{equation}
Here $\prod_{e \in v} X_e$ denotes the product of six Pauli-X operators on the edges emanating from the vertex $v$ [see Fig.1(c) in the main text] and $g$ denotes all possible closed surfaces on the dual lattice.
When subjecting the initial pure state $\rho_0= |\Psi_0\rangle \langle \Psi_0|$ to the phase-flip channel $\mathcal{E}_e [\cdot] = (1-p)\rho  + p Z_e \rho Z_e$ , one finds 
\begin{equation}
	\label{Eq:3dtoric_loop}
	\rho = \sum_{g,\bar{g} } \frac{e^{-K|g \bar{g}|}}{N_G} |g\rangle \langle \bar{g} |,
\end{equation}
where $K = -\log(1-2p)$, $N_G$ denotes the number of  possible closed surface configurations, and $|g \bar{g}|$ denotes the total area of the closed surfaces $g\bar{g}$.
To compute the 2n-th Renyi negativity, let's first investigate what statistical model $\tr(\rho^{2n})$ is mapped to using Eq.\eqref{Eq:3dtoric_loop}.
Before doing the calculation, we first note that since we have already shown in the main text that $\rho = \sum_{x_\mathbf{e}} \mathcal{Z}_{\text{3d gauge}, x_e} |\Omega_{x_\mathbf{e}} \rangle \langle \Omega_{x_\mathbf{e}}|$, one must have $\tr(\rho^{2n}) = \sum_{x_\mathbf{e}} (\mathcal{Z}_{\text{3d gauge}, x_e})^{2n}$.
However, it will be clear soon that expressing $\tr(\rho^{2n})$ using Eq.\eqref{Eq:3dtoric_loop} will provide a more intuitive understanding about the subleading $\log 2$ corrections of the Renyi negativity when $p< p_c^{(2n)}$.
To proceed, a simple use of $\langle g | g'\rangle = \delta_{g,g'}$ yields
\begin{equation}
	\label{Eq:3d_rho2n}
	\tr(\rho^{2n}) = \frac{1}{(N_G)^{2n}} \sum_{g^{(s)}} e^{-K \sum_{s = 1}^{2n} |g^{(s)} g^{(s+1)_{2n}}|}.
\end{equation}
where $(s+1)_{2n}$ denotes the sum modulo $2n$.
Eq.\eqref{Eq:3d_rho2n}  can be regarded as  $2n$ different flavors of closed surface models with the $s$-th flavored closed surface coupled to the $(s+1)_{2n}$-th flavored one through their relative closed surface configurations $|g^{(s)} g^{(s+1)_{2n}}|$.
One can also map it to the 3d spin model by identifying the $s$-th flavored closed surface as the domain wall of the $s$-th flavored Ising spins, and Eq.\eqref{Eq:3d_rho2n} can be written as $\tr(\rho^{2n})  \propto  \sum_{ \{\sigma^{(s)}_\mathbf{v} \} } e^{-H_{2n}}$,
where
\begin{equation}
	\label{Eq:spin_chamon_n}
	-H_{2n}  = \frac{K}{2}  \sum_{ \langle v,v'\rangle }\sum_{s = 1}^{2n}      \sigma^{(s)}_{v} \sigma^{(s)}_{v'}   \sigma^{(s+1)_{2n}}_{v} \sigma^{(s+1)_{2n}}_{v'} .
\end{equation}
We can therefore regard the $2n$-th moment density matrix of the decohered 3d toric code as the $2n$-flavored 3d Ising model with the $s$-th flavored  spin coupled to the $(s+1)_{2n}$-th flavored one through the four-spin interactions.
One expects that there is a critical inverse temperature $K_{c}^{(2n)}$ such that when $K$ is above(below) $K_c^{(2n)}$, the relative spins $\sigma^{(s)}_v \sigma^{(t)}_v, \forall s, t$ is ordered (disordered).
We note that the relation $K = -\log(1-2p)$ implies the low-error limit corresponds to the high-temperature limit, which is opposite to the mapping in the main text where low-error limit is mapped to the low-temperature limit through $\tanh(\beta) = 1-2p$.
Besides, we get the Ising spin models (Eq.\ref{Eq:spin_chamon_n}) instead of the Ising gauge model.
All these are common features of standard Kramer-Wannier type dualities.


We now compute $\tr[(\rho^{T_A})^{2n}]$, the numerator in Eq.\eqref{Eq:renyi_negativity_def}.
Denoting $\rho_{g, \bar{g}} = e^{-K |g \bar{g}|}/N_G$, the partial transpose of the density matrix then takes the form
\begin{equation}
	\rho^{T_A} = \sum_{g,\bar{g} } \rho_{g,\bar{g}} |\bar{g}_A , g_B\rangle \langle g_A , \bar{g}_B |.
\end{equation}
It follows that the $n$-th moment of $\rho^{T_A} $ can be evaluated as
\begin{equation}
	\label{Eq:tr_rho_TA}
	\begin{aligned}
		\tr[(\rho^{T_A})^{2n}] & = \sum_{ \{ g^{(s)} \}, \{ \bar{g}^{(s)} \} } \Big(\prod_{s = 1}^n \rho_{g^{(s)}, \bar{g}^{(s)}}
		\Big) 
		\Big( \prod_{s = 1}^n \langle g^{(s)}_{A},  \bar{g}^{(s)}_{B}| \bar{g}^{(s+1)}_{A},  g^{(s+1)}_{B}\rangle \Big)\\
		& = \sum_{ \{ g^{(s)} \}, \{ \bar{g}^{(s)} \} } 
		\Big(\prod_{s = 1}^n \rho_{g^{(s)}, \bar{g}^{(s)} g^{(s+1)}}\Big)
		\Big(\prod_{s = 1}^n \langle g^{(s)}_{A},  \bar{g}^{(s)}_{B} {g}^{(s+1)}_{B} | \bar{g}^{(s+1)}_{A} {g}^{(s+2)}_{A},  g^{(s+1)}_{B}\rangle \Big),
	\end{aligned}
\end{equation}
where we have relabeled the dummy variables $\{\bar{g}^{(s)} \}$ as $\{\bar{g}^{(s)} {g}^{(s+1)} \}$ in the second line.
To simplify Eq.\eqref{Eq:tr_rho_TA}, we first note that the inner product enforces the constraints  $\bar{g}^{(s)}_{B} = I$ and $g_A^{(s)} = \bar{g}^{(s+1)}_{A} {g}^{(s+2)}_{A}$.
The constraint $\bar{g}^{(s)}_{B} = I$ implies $\bar{g}^{(s)} = \bar{g}^{(s)}_A \in G_A$ is restricted to closed surfaces \textit{merely} in region $A$. 
Therefore,
\begin{equation}
	\begin{aligned}
		\tr[(\rho^{T_A})^{2n}] & = \sum_{ \{ g^{(s)} \}, \{ \bar{g}^{(s)}_A \in G_A \} } 
		\Big(\prod_{s = 1}^n \rho_{g^{(s)}, \bar{g}_A^{(s)} g^{(s+1)}_A g^{(s+1)}_B}\Big)
		\Big(\prod_{s = 1}^n \langle g^{(s)}_{A}| \bar{g}^{(s+1)}_{A} {g}^{(s+2)}_{A}\rangle \Big) \\
		& =  \sum_{ \{ g^{(s)} \}} 
		\Big(\prod_{s = 1}^n \rho_{g^{(s)}, g_A^{(s-1)} g^{(s+1)}_B}\Big) 
		\sum_{\{ \bar{g}^{(s)}_A \in G_A \} }\Big(\prod_{s = 1}^n \langle g^{(s)}_{A}| \bar{g}^{(s+1)}_{A} {g}^{(s+2)}_{A}\rangle \Big).
	\end{aligned}
\end{equation}
In the second line, we use $g_A^{(s)} = \bar{g}^{(s+1)}_{A} {g}^{(s+2)}_{A}$ (the constraint enforced by the inner product) to rewite $\prod_{s = 1}^n \rho_{g^{(s)}, \bar{g}_A^{(s)} g^{(s+1)}_A g^{(s+1)}_B}$ so that it is independent of $\bar{g}^{(s)}$.
Let's now consider the product $\prod_{s = 1}^n \rho_{g^{(s)}, g_A^{(s-1)} g^{(s+1)}_B}$  and the sum of the products $\sum_{\{ \bar{g}^{(s)}_A \in G_A \} }\Big(\prod_{s = 1}^n \langle g^{(s)}_{A}| \bar{g}^{(s+1)}_{A} {g}^{(s+2)}_{A}\rangle \Big)$ separately.
Remarkably, the former takes a simple form
\begin{equation}
	\begin{aligned}
		\prod_{s = 1}^n \rho_{g^{(s)}, g_A^{(s-1)} g^{(s+1)}_B} & = \frac{1}{(\mathcal{Z}N_G)^{2n}}e^{- K \sum_s (|g_A^{(s)} g_A^{(s-1)}| + |g_B^{(s)}g_B^{(s+1)}|)} \\
		& = \frac{1}{(\mathcal{Z}N_G)^{2n}}e^{ - K \sum_s |g^{(s)} g^{(s-1)}|} = \frac{e^{-H_{2n}}}{(\mathcal{Z}N_G)^{2n}}.
	\end{aligned}
\end{equation}
On the other hand, the requirement that $\prod_{s = 1}^n \langle g^{(s)}_{A}| \bar{g}^{(s+1)}_{A} {g}^{(s+2)}_{A}\rangle $ is non-vanishing enforces the constraint that $g^{(s+2)}$ and $ g^{(s)}$ can only differ from closed curfaces in region $A$ and region $B$ but not the surfaces that cross the boundary between $A$ and $B$.
It follows that $g^{(2s+1)} $ with $s = 1,...,n-1$ are related to $g^{(1)}$ through $g^{(2s+1)} = h_A^{(2s-1)} g^{(1)}h_B^{(2s-1)}$ with $h_A^{(2s-1)} \in G_A$, $h_B^{(2s-1)} \in G_B$ , and similarly $g^{(2s+2)} $ with $s = 1,...,n-1$ are related to $g^{(2)}$ through $g^{(2s+2)} = h_A^{(2s)} g^{(2)}h_B^{(2s)}$.
Therefore, there are $2n-2$ constraints in the summation of $\sum_{\{g^{(s)}\}}$, which we denote as $\sum_{\{g^{(s)}\}_{A} }$.
The logarithmic Renyi negativity can then be written as
\begin{equation}
	\mathcal{N}^{(2n)}_A  = \frac{1}{2-2n}\Bigg[ \log \Big(\sum_{\{g^{(s)}\}_A }e^{-H_{2n}[\{g^{(s)}\}]} \Big) - \log \Big(\sum_{\{g^{(s)}\} }e^{-H_{2n}[\{g^{(s)}\}]} \Big) \Bigg] = \frac{F_A^{(2n)} - F_0^{(2n)}}{2-2n},
\end{equation}
which is the excess free energy of enforcing the aforementioned constraints.
These constraints can be understood intuitively by again interpreting the closed surfaces as the domain walls of the Ising spins. 
The requirement that $g^{(2s+1)} = h_A^{(2s-1)} g^{(1)}h_B^{(2s-1)}$ with $h_A^{(2s-1)} \in G_A$, $h_B^{(2s-1)} \in G_B$ then corresponds to the constraint that all  odd-flavored Ising spins are forced to have the same domain-wall configurations as the 1st-flavored Ising spins on the boundary of region $A$ (similar relations hold for the even-flavored spins).
In the paramagnetic phase $K < K^{(2n)}_c$, since the aligned boundary spins can fluctuate together without changing the domain wall configurations, one finds the subleading $\log 2 $ correction, which we identify as the topological negativity. 
On the other hand, such degrees of freedoms are absent in the ferromagnetic phase, as the relative spins $\sigma^{(s)}_v \sigma^{(t)}_v, \forall s, t$ are forced to order together, leading to the vanishing of the topological negativity.
\section{3d toric code with bit-flip errors}
\label{sec:3dtoricbitflip}
Here we discuss separability transition induced by bit-flip errors in the 3d toric code. Following the general strategy discussed in the main text, the parent cluster state should be such so that subjecting the ground state density matrix of the 3d toric code to bit-flip errors (Kraus operators $\sim X_e$) results in a Gibbs form. Therefore, we consider $H'_{\textrm{3d Cluster}}  = -\sum_v Z_v (\prod_{e \ni v} X_e) - \sum_e Z_e (\prod_{v \in e} X_v)$, which has previously also appeared in Ref.\cite{lee2022measurement}. The rest of the analysis is quite similar to that for the 2d toric code (with $X \leftrightarrow Z$ everywhere). After writing the non-decohered density matrix of toric code as $\rho_{0} \propto \langle z_\mathbf{v} = 1| \rho_{C,0} |z_\mathbf{v} = 1\rangle$, where $\rho_{C,0} $ is the ground state of $H'_{\textrm{3d Cluster}} $, and subjecting it to bit-flip errors, the decohered state is schematically given by 	$\rho \propto \sum_{z_\mathbf{e}}   \mathcal{Z}_{\text{3d Ising}, z_\mathbf{e}}  |\Omega_{z_\mathbf{e}} \rangle \langle \Omega_{z_\mathbf{e}} | $, where $|\Omega_{z_\mathbf{e}}\rangle$ are toric code eigenstates and $\mathcal{Z}_{\text{3d Ising}, z_\mathbf{e}}$   is the partition function of the 3d classical Ising model with interactions determined by $\{z_e\}$. The analog of the state $|\psi(\beta)\rangle$ is $|\psi^{} (\beta) \rangle \propto  \sum_{x_\mathbf{e}} [\mathcal{Z}_{\text{3d Ising}, z_\mathbf{e}} (\beta)]^{1/2} |z_\mathbf{e} \rangle$ and its topological transition is indicated by the non-analyticity of the `t Hooft operator similar to the discussion of the 2d toric code. The corresponding $p_c$ for the separability transition is then determined by the transition out of the ferromagnetic phase in the 3d RBIM along the Nishimori line, which matches the optimal error-recovery threshold, $p_c \approx 0.233$ \cite{ozeki1998multicritical, kubica2018three}.

\section{Entanglement negativity of $\rho^{(2)} \equiv \rho^2/\tr(\rho^2)$ for 2d toric code}
\label{sec:negativity}

The goal of this appendix is to show that the topological entanglement negativity of $\rho^{(2)} \equiv \rho^2/\tr(\rho^2)$ vanishes at $p^{}_c = (1-e^{-K_c})/2 = (1-\sqrt{\sqrt
2 - 1})/2$, where $K_c$ is the critical inverse temperature of the translationally invariant 2d classical Ising model.
We will compute the $2n$-th Renyi entanglement negativity defined as:
\begin{equation}
	\label{Eq:renyi_negativity}
\mathcal{N}^{2n}_A(\rho^{(2)}) = \frac{1}{2-2n} \log \frac{ \tr \{[(\rho^2)^{T_A}]^{2n} \}  }{\tr[( \rho^{2})^{2n} ] }.
\end{equation}
%
Since the calculation is almost the same as the one in Ref.\cite{fan2023diagnostics}, we will only breifly sketch the derivation and refer to Ref.\cite{fan2023diagnostics} for details.

It is more convenient to use the `loop representation' of $\rho$ mentioned in Ref.\cite{fan2023diagnostics} :
\begin{equation}
	\label{Eq:toric_loop}
	\rho \propto \sum_{g_x } e^{-K |g_x|} g_x \sum_{g_z } g_z.
\end{equation}
Here $e^{-K} = \tanh(\beta) = (1-2p)$, $\gamma$ denotes all possible loop configurations, $g_x$ denotes the Pauli-X loops, and $|g_x|$ denotes the total lengths of $g_x$. The relation for $g_z$ is similar.
We note that Eq.\eqref{Eq:toric_loop}  can also be easily obtained from Eq.(1) of the main text, i.e., $\rho_T \propto \langle  x_\mathbf{v} = 1|
e^{-\beta \sum_e h_{e}}	| x_\mathbf{v} = 1\rangle \prod_v (I + \prod_{e \ni v} Z_e) $.
To see this, we note that $\prod_v[ (I + \prod_{e \ni v} Z_e)] = \sum_{g_z \in \gamma} g_z$ is already in the loop representation.
A simple expansion $e^{-\beta h_e} \propto   \cosh(\beta)I + \sinh(\beta) X_e( \prod_{v \in e} Z_v)$ and the fact that $\langle x_v = 1 |Z_v |x_v = 1\rangle = 0$ then yield  Eq.\eqref{Eq:toric_loop}.
Now, $\rho^{(2)}$ can be written as 
\begin{equation}
\begin{aligned}
\rho^{(2)} & \propto \sum_{g_x, h_x} e^{-K (|g_x| + |h_x|) } g_x h_x  \sum_{g_z, h_z} g_z h_z\\
 & \propto \sum_{g_x, g_z} (\sum_{h_x} e^{-K (|g_x h_x| + |h_x|) } ) g_x  g_z.
\end{aligned}
\end{equation}
In the second line, we relabeled the dummy variable $g_x$ and $g_z$ as $g_x h_x$ and $g_z h_z$, respectively.
Noting that the trace of the product of Pauli operators is non-vanishing only if it is proportional to identity, one finds $\tr[(\rho^{(2)})^{2n}] $  is proportional $\mathcal{Z}^{(4n)}(K)$, the partiton function of the $(4n-1)$-flavored Ising model with inverse temperature $K$ mentioned in Ref.\cite{fan2023diagnostics}.
On the other hand, to compute $(\rho^{(2)})^{T_A}$, we note that 
\begin{equation}
(g_x g_z)^{T_A} = (-1)^{\#\textrm{cross}_A (g_x , g_z)} g_x  g_z,
\end{equation}
where $\#\textrm{cross}_A (g_x , g_z)$ denotes the number of crossings between $g_x$ and $g_z$ in region $A$.
Therefore,
\begin{equation}
	\begin{aligned}
	(\rho^{(2)})^{T_A} &  \propto \sum_{g_x, g_z} (\sum_{h_x} e^{-K (|g_x h_x| + |h_x|) } )  (-1)^{\#\textrm{cross}_A (g_x , g_z)} g_x  g_z.
	\end{aligned}
\end{equation}
A similar derivation in Ref.\cite{fan2023diagnostics} then shows that $\tr \{[(\rho^2)^{T_A}]^{2n} \}$ is proportional to the partiton function of the $(4n-1)$-flavored Ising model with $(2n-2)$ constraints that force the $|\partial A|$ boundary spins aligning in the same direction.
Therefore, the $2n$-th Renyi entanglement negativity in Eq.\eqref{Eq:renyi_negativity} corresponds to the excess free energy for aligning one species of Ising spins for the $(4n-1)$-flavored Ising model.
As $2n\rightarrow 1$, $\mathcal{Z}_{4n} \rightarrow \mathcal{Z}_{2}$ becomes the partition function of the 2d Ising model \cite{fan2023diagnostics}, and thus the topological entanglement negativity vanishes at exactly the same $p_c$ corresponding to the translationally invariant Ising model.

\end{document}